\documentclass[12pt]{article}

\usepackage{newtxtext,newtxmath}
\usepackage{ccaption}

\usepackage{graphicx}
\usepackage{aas_macros}

\usepackage[letterpaper,margin=1in]{geometry}

\renewenvironment{abstract}
	{\quotation}
	{\endquotation}

\date{}

\newcommand{\Msun}{M$_\odot$\,}
\newcommand{\Lsun}{L$_\odot$}
\newcommand{\Rsun}{R$_\odot$}

\newcommand\arcmin{\mbox{$^\prime$}}%
\newcommand\arcsec{\mbox{$^{\prime\prime}$}}%
\makeatletter
\renewcommand{\fnum@figure}{\textbf{Figure \thefigure}}
\renewcommand{\fnum@table}{\textbf{Table \thetable}}
\makeatother

\usepackage{scicite}

\usepackage{url}

\def\scititle{
	Disappearance of a massive star in the Andromeda Galaxy due to formation of a black hole}
\title{\bfseries \boldmath \scititle}

\author{Kishalay De$^{1,2,3\ast}$, Morgan MacLeod$^{4}$, Jacob E. Jencson$^{5}$, Elizabeth Lovegrove$^6$,\and
Andrea Antoni$^2$, Erin Kara$^{3}$, Mansi M. Kasliwal$^{7}$, Ryan M. Lau$^{8}$, Abraham Loeb$^{4,9}$,\and
Megan Masterson$^3$, Aaron M. Meisner$^{8}$, Christos Panagiotou$^3$, Eliot Quataert$^{10}$\and
and Robert Simcoe$^{3}$\and
\small$^{1}$ Department of Astronomy and Columbia Astrophysics Laboratory, Columbia University, New York, USA \and
\small$^{2}$ Center for Computational Astrophysics, Flatiron Institute, New York, USA \and
\small$^{3}$ Kavli Institute for Astrophysics and Space Research, Massachusetts Institute of Technology, Cambridge, USA\and
\small$^{4}$ Center for Astrophysics, Harvard University \& Smithsonian, Cambridge, USA\and
\small$^{5}$ IPAC, Caltech, Pasadena, USA \and
\small$^{6}$ United States Naval Observatory, Washington DC, USA\and
\small$^{7}$ Cahill Center for Astrophysics, California Institute of Technology, Pasadena, USA\and
\small$^{8}$ National Optical-Infrared Astronomy Research Laboratory, National Science Foundation, Tucson, USA\and
\small$^{9}$ Black Hole Initiative, Harvard University, Cambridge, USA\and
\small$^{10}$ Department of Astrophysical Sciences, Princeton University, Princeton, USA\and
\small$^\ast$Corresponding author. Email: kd3038@columbia.edu
}

\begin{document} 

\maketitle

\begin{abstract} \bfseries \boldmath
When a massive star reaches the end of its lifetime, its core collapses and releases neutrinos that drive a shock into the outer layers (stellar envelope). A sufficiently strong shock ejects the envelope, producing a supernova. If the shock fails to eject it, the envelope is predicted to fall back onto the collapsing core, producing a stellar-mass black hole (BH) and causing the star to disappear. We report observations of M31-2014-DS1, a hydrogen-depleted supergiant in the Andromeda Galaxy. In 2014 it brightened in the mid-infrared. From 2017 to 2022 it faded by factors of $\gtrsim10^4$ in optical light, becoming undetectable, and $\gtrsim10$ in total light. We interpret these observations, and those of a previous event in NGC\,6946, as evidence for failed supernovae forming stellar-mass BHs.
\end{abstract}

Massive stars [those with initial masses $\gtrsim 10$\,solar masses (\Msun)] can expand and become variable towards the end of their lifetime, undergoing luminosity changes observable on human timescales \cite{Langer2012}. These variations can result from unstable nuclear fusion in the core or mass transfer to a binary companion as the star expands \cite{Smith2011}. Some of these stars end their lives in luminous [$\gtrsim 10^7$\,solar luminosities (L$_\odot$)] supernova (SN) explosions caused by the collapse of the core \cite{Janka2007, Burrows2021} and subsequent ejection of the stellar envelope by neutrino-powered shocks \cite{Langer2012, Smartt2009}. These core-collapse SNe are routinely observed by optical time-domain surveys \cite{Perley2020}.

Theoretical models predict that in other massive stars, the neutrino shock fails to eject the envelope, causing most of the stellar material to fall back onto the collapsing core -- forming a stellar mass black hole (BH) \cite{Kochanek2008, Smartt2015}. Even as the star abruptly disappears following the failed SN \cite{OConnor2011, Smartt2009}, the initial loss of gravitational binding energy (equivalent to $\approx 0.2$ to $0.5$\,\Msun) to neutrino emission \cite{Nadezhin1980, Lovegrove2013} and feedback from inefficient accretion \cite{2016ApJ...827...40G, Quataert2019} are predicted to inject $\sim 10^{45}$ to $10^{49}$\,erg into the stellar envelope. This is sufficient to eject a small fraction of the outer envelope, possibly producing a faint optical transient ($\lesssim 10^6\,$L$_\odot$; \cite{Lovegrove2013, Lovegrove2017}). A long-lived brightening in mid-infrared (MIR) light can result from the condensation of this material in a circumstellar dust shell \cite{Kochanek2014, Adams2017}. Observing such low luminosity stellar eruptions and stars that disappear without an explosion requires monitoring individual stars at extragalactic distances \cite{Reynolds2015, Neustadt2021} with uniform, sensitive searches over years to decades \cite{Kochanek2008}.

\subsection*{Observations of a disappearing supergiant}
\begin{figure}[!ht]
    \centering
    \includegraphics[width=0.99\textwidth]{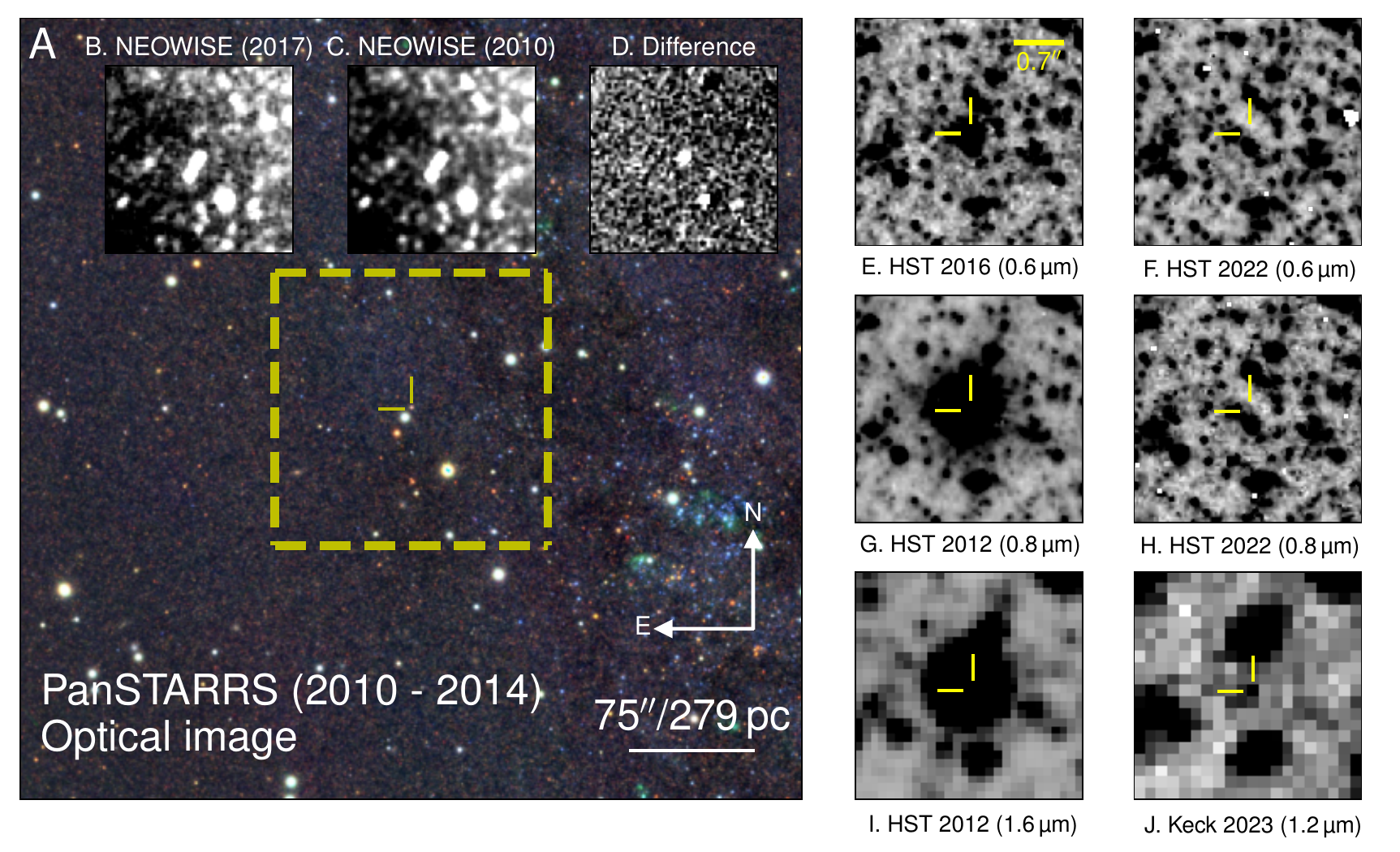}
    \linespread{1.0}\selectfont{}
    \caption{\textbf{Location and disappearance of M31-2014-DS1.} (A) Optical color composite image of the discovery field taken from the Panoramic Survey Telescope and Rapid Response System (PanSTARRS/PS1 survey; \cite{suppmat}). The yellow dashed square indicates the region shown in panels B-D (where white indicates brighter pixels) and the yellow cross-hair marks the position of the star. (B) NEOWISE\cite{Mainzer2014} MIR image taken in 2017, (C) NEOWISE image in 2010 and (D) the difference between them. The other panels show zoomed-in images of the star (as indicated by the scale, with black indicating brighter pixels) taken in the labeled years: (E-H) optical HST images; (I) near-infrared HST image; (J) near-infrared Keck image.}
    \label{fig:cutout}
\end{figure}

We applied an image subtraction pipeline\cite{De2023} to the Near-Earth Object Wide-Field Infrared Survey Explorer (NEOWISE; \cite{Mainzer2014}) MIR sky survey to search for variable sources in the Andromeda and Triangulum local group galaxies cataloged as Messier 31 and Messier 33, respectively. Using the six-month cadenced observations from 2009 to 2022, we searched for luminous MIR transients \cite{suppmat} that would accompany dusty stellar eruptions such as failed SNe. We identified a faint brightening of a star (hereafter M31-2014-DS1) at celestial coordinates Right Ascension 00h 45m 13.47s, Declination +41$^\circ$ 32$'$ 33.14$''$ (J2000 equinox) (Figures \ref{fig:cutout}B-D) towards M31. Beginning in 2014, this source source increased in MIR flux by $\approx 50$\% over $\approx 2$\,years, then faded below its initial flux within a year, and continued to fade until 2022 (Figures \ref{fig:optirlc}, \ref{fig:prog} \& \ref{fig:linearlc}).

We retrieved optical light curves of the source from ground and space-based surveys \cite{suppmat}. In optical light, this source faded by a factor of $\gtrsim 100$ between 2016 and 2019, and was undetectable in recent ground-based optical imaging using the MMT Observatory in 2023 (Figure \ref{fig:optirlc}). This location was also serendipitously imaged by the Hubble Space Telescope (HST) in 2022. We re-analyzed the HST data \cite{suppmat}, finding no detection in the optical F606W filter (Figure \ref{fig:cutout}F), but a faint source ($\gtrsim 10^4\times$ fainter than the progenitor from 2012) in the near-infrared (NIR) F814W filter (Figure \ref{fig:cutout}H). We obtained follow-up NIR imaging and spectroscopic observations using the Infrared Telescope Facility and Keck telescopes in 2023 \cite{suppmat}, which confirm a faint red source in the NIR $JHK$-bands (Figures \ref{fig:cutout}J and \ref{fig:evol_remnant}B).

This source was previously identified as an irregular variable star (designated V7984\,M31B; \cite{Kaluzny1998}), and classified as a candidate red supergiant (RSG) based on archival NIR colors \cite{Massey2021}. We used archival HST and Spitzer Space Telescope (SST) observations from 2005-2012 \cite{suppmat} to measure the star's spectral energy distribution (SED). We fitted the SED with a model of a blackbody photosphere surrounded by a circumstellar dust shell expected from mass loss in evolved stars (Figure \ref{fig:prog}). The best-fitting model corresponds to a supergiant star with luminosity $L\approx10^5 L_\odot$ and effective temperature $T_{\rm eff} \approx 4500$\,K surrounded by a dust shell with temperature $T_d\approx 870$\,K at a radius of $\approx 110$\,astronomical units (au). Although previously classified as a candidate RSG, this effective temperature classifies the star as a warmer, yellow supergiant \cite{Beasor2023b, Humphreys2023}. The observed red SED is due to reddening by the circumstellar dust, from which we infer a high mass-loss rate of $\approx 10^{-4}$\,\Msun\,yr$^{-1}$\cite{suppmat}.

To investigate how these properties varied over time, we applied the same SED model to optical data from the Gaia space telescope and MIR data from NEOWISE \cite{suppmat}. The best-fitting models have an almost constant bolometric luminosity (radiated power integrated across all wavelengths) for $\approx 1000$\,days after the MIR brightening began, which then declines by $\gtrsim 10\times$ over the next $\approx 1000$\,days \cite{suppmat} (Figure \ref{fig:model}). During this time, the source became more obscured by dust (Figures \ref{fig:evol_remnant} \& \ref{fig:dust_param_evol}), with the dust shell reaching an optical depth $\tau \gtrsim 20$ and the central blackbody reducing in radius by a factor of $\gtrsim 5$ \cite{suppmat} (Figure \ref{fig:evol_remnant}).

\subsection*{Core-collapse of a massive star}

The abrupt and sustained bolometric fading of M31-2014-DS1 is unlike the variations observed in other massive, evolved stars\cite{Conroy2018, Soraisam2020}. Some sources exhibit a temporary optical dimming due to episodes of enhanced mass loss and dust formation \cite{2022ApJ...930...81J, 2022ApJ...936...18D}, but this re-processes the optical emission into the infrared, so the bolometric luminosity remains constant or gets brighter \cite{suppmat} due to ongoing nuclear fusion and sometimes energy injection from a binary companion \cite{Kochanek2011, Smith2011b}. Even with dust obscuration, those sources become brighter in the infrared. Geometric effects, such as preferential mass loss along the equator \cite{Kashi2017} can reduce the fraction of the remnant luminosity that reaches an observer; however, calculations show this can reduce the observed luminosity by no more than a factor of two \cite{Kochanek2024}. This expectation is consistent with observations of nearby stellar merger remnants with known mass ejection geometry \cite{Kochanek2024}.

The continuous optical and MIR observations of M31-2014-DS1 during its fading (Figure \ref{fig:optirlc}) show no commensurate increase in infrared brightness while the optical brightness faded by a factor of $10^4$. Therefore the bolometric luminosity reduced, which we attribute to a cessation in nuclear fusion and the collapse of the stellar core. There is no evidence for an associated core-collapse SN, which would have been easily observable. We therefore interpret M31-2014-DS1 as a failed SN in which most of the stellar envelope fell back to form a BH \cite{Kochanek2008}.

\subsection*{Interpretation as a failed supernova}

We compare the inferred properties of M31-2014-DS1 to stellar evolution models and the theoretically expected end-stages of massive stars. The observed progenitor properties (Figure \ref{fig:prog}, between 2005 to 2012) are hotter than predicted for the nominal end-point of single star evolutionary tracks \cite{Choi2016} which characterize the progenitors of the hydrogen-rich SNe\,IIP; however, they are similar to the observed progenitors of some hydrogen-deficient SNe (the Type IIb and IIL SNe) \cite{Smartt2009}. The latter have been interpreted as arising from high mass loss (due to winds or binary interaction; \cite{Georgy2012, Yoon2010}) that removed most of the hydrogen-rich stellar envelope. We constructed stellar models of that process \cite{suppmat} and find the closest match with the observed progenitor has a terminal mass of $\approx 5$\,\Msun, of which $\approx 0.28$\,\Msun is the outer hydrogen-rich envelope (Figure \ref{fig:prog}B). This model had an initial mass of $\approx 13$\Msun, and is bluer at late times than a hydrogen-rich counterpart without enhanced mass loss \cite{suppmat}.

We use the optical observations to empirically constrain any mass ejection associated with the collapsing stellar core. Based on analytical models for outbursts caused by energy injection into hydrogen-rich envelopes \cite{Kasen2009}, we set an upper limit of $\lesssim 0.1$\,\Msun ejected material at the star's escape velocity $v_{\rm esc}\approx 60$\,km\,s$^{-1}$, and corresponding limits of $\lesssim 0.3$\,\Msun\ at $5\,v_{\rm esc}$ and $\lesssim 1.5$\,\Msun\ at $25\,v_{\rm esc}$ (Figure \ref{fig:model}A). These constraints imply that most of the $\approx 5$\,\Msun\ star collapsed instead of being ejected, exceeding the maximum mass of a neutron star \cite{Kalogera1996} and therefore forming a BH.

If the star were collapsing purely due to its self-gravity, the outer envelope would collapse into the BH within roughly its free-fall time $t_{\rm ff} \approx \frac{1}{\sqrt{G\rho}} \approx 210$\,d, where $G$ is the Gravitational constant and $\rho$ is the average stellar density. This is shorter than the observed continued bolometric fading over $\gtrsim 1000$\,d (Figure \ref{fig:model}), indicating that the fallback is delayed by energy injected into the outer envelope. To simulate the resulting mass accretion onto the central BH, we take the same hydrogen-depleted progenitor star model as above but add a range of shock energies ($\sim 10^{45}$ to $10^{49}$\,erg) into the outer envelope. Such shocks are expected from neutrino mass loss and feedback from inefficient accretion \cite{Quataert2012,2021ApJ...911....6I, Antoni2023} during the collapse of the core. In the final stages of stellar evolution, these outer layers undergo vigorous turbulent convection, which carries substantial angular momentum \cite{2016ApJ...827...40G,Quataert2019} and suppresses direct accretion \cite{1986ApJ...308..755B}. Numerical simulations\cite{Antoni2023} show that $\lesssim 1$\% of material falling from the outer supergiant envelope accretes directly onto the BH. We adopt an analytic model of the accreted fraction as a function of the relative angular momenta of the fallback material and that of an orbit at the BH horizon \cite{suppmat}.

The results from our models indicate that the mass accretion rate declines rapidly at early times, initially far exceeding the maximum accretion rate (the Eddington accretion rate) where photons can freely escape from near the BH \cite{Jiang2024}. During this phase, the emergent luminosity is expected \cite{Jiang2014} to be capped near the Eddington luminosity $L_{\rm Edd} \approx 6 \times 10^{38}$\,erg\,s$^{-1}$ (for a $5$\,M$_\odot$ BH), corresponding to the limit where material accreting at the Eddington accretion rate generates accretion luminosity such that the outward radiation pressure balances the inward gravitational pull of the BH.

Comparing these models to the observations (Figure \ref{fig:model}) indicates that the $\approx 1000$\,d luminosity plateau is explained by a super-Eddington accretion phase when the source luminosity remains steady at $\approx 30$ to $50$\% $L_{\rm Edd}$ despite more rapid mass infall. As the infall rate falls below the Eddington rate, the emergent luminosity is expected to become proportional to the decaying accretion rate, as observed in the data. Larger shock energies unbind a larger fraction of the progenitor envelope, leaving behind a smaller mass that falls back over a shorter period of time. Although we cannot precisely constrain the exact time of collapse, the fading by a factor of $\approx 10$ over $\approx 1000$\,d is most consistent with shock energies of $10^{47}$ to $10^{48}$\,erg (Figure \ref{fig:model}). In those cases, $\approx 98$\% of the stellar mass collapses or falls back, leaving a $\approx 5M_\odot$ BH.

Those are weak shock energies compared to typical core collapse SNe ($\sim 10^{51}$\,erg; \cite{Smartt2009}). They are sufficient to unbind $\lesssim 0.1 M_\odot$ of the outer envelope, which would produce a luminous optical outburst that is sufficiently brief to have been missed by observations given the photometric cadence (Figure \ref{fig:model}). We expect the ejected mass to carry a distribution of velocities as it moves outwards and cools \cite{suppmat}. Most of the material would reach the dust condensation radius ($r_{\rm c}$) at which solid dust grains begin to condense from the ejecta. This would produce hot dust on a timescale of $\sim r_{\rm c} / v_{\rm esc} \approx 30\,{\rm au}/ 60\,{\rm km\,s}^{-1} \approx 900$\,d after the ejection, which is consistent with the time of the observed peak MIR flux. Although the details of the dust formation are sensitive to the velocity distribution and dust formation process\cite{Kochanek2011}, our model velocity distributions \cite{suppmat} imply that the fraction of the total ejecta in the hot dust shell (near $r_{\rm c}$) is $f_{\rm shell} \approx 0.1$ at any given time. The hot dust mass in the shell is therefore $m_{\rm shell} \sim f_{\rm shell}  M_g  r_{\rm dg}$, where $M_g$ is the total ejected gas mass ($\sim 0.1$\,\Msun) and $r_{\rm dg} \approx 0.01$ \cite{Karambelkar2025, Kaminski2021} is the assumed dust-to-gas mass ratio. 
We find $m_{\rm shell} \sim 10^{-4}$\,\Msun, which is similar to the dust mass derived from the SED ($\approx 1.1 \times 10^{-4}$\,\Msun; Figure \ref{fig:dust_param_evol}).

\subsection*{Unified model for a previous candidate}

A candidate disappearing supergiant (designated NGC\,6946-BH1) has previously been identified in the galaxy NGC\,6946\cite{Gerke2015,Adams2017,Basinger2021}. Figure \ref{fig:prog} also shows its progenitor for comparison with M31-2014-DS1. NGC\,6946-BH1 exhibited a luminous ($\sim 10^6$\,\Lsun) optical outburst followed by an expanding dusty envelope, implying $\sim 0.1 -1$\,\Msun of ejecta \cite{Basinger2021}. This was consistent with predictions of either failed SNe from hydrogen-rich progenitors \cite{Lovegrove2013} or some types of stellar mergers \cite{Kashi2017, Beasor2023}. The dust-obscured remnant then faded to $\approx 15$\% ($\sim 10^{38}$\,erg\,s$^{-1}$) of the progenitor luminosity in $\approx 3000$\,days \cite{Adams2017}, indicating the termination of nuclear fusion \cite{Kochanek2023}. However, this is difficult to reconcile with the decades-long ($\gtrsim 10^5$\,d) super-Eddington phase for massive hydrogen-rich stars \cite{Fernandez2018, Basinger2021}, during which the emergent luminosity is expected to be constant near the BH Eddington limit ($\sim 10^{39}$\,erg\,s$^{-1}$). By constructing a hydrogen-rich RSG progenitor model \cite{suppmat} that matches the M31-2014-DS1 progenitor luminosity, we confirm that rapid fading timescales of a few years cannot be explained with stars that have largely retained their hydrogen-rich envelopes (Figure \ref{fig:model}).

The progenitor of NGC\,6946-BH1 was initially identified as an RSG\cite{Gerke2015, Adams2017}. Later analysis found that it was hotter than expected for single star RSG evolution (Figure \ref{fig:prog}), being more consistent with a yellow supergiant \cite{Humphreys2019}. We compare its properties to our models, finding that the closest match is a progenitor with an initial mass of $17.5M_\odot$ with enhanced late-stage mass loss, which forms a hydrogen-depleted star with a terminal mass of $\approx 7.5$\,\Msun of which $\approx 0.6$\,\Msun is the hydrogen-rich envelope (Figure \ref{fig:prog}; \cite{suppmat}). Applying a similar analysis as for M31-2014-DS1 \cite{suppmat}, we calculate that the twice as massive hydrogen-rich envelope in the NGC 6946-BH1 model produces a longer duration outburst ($\sim 150$\,d), as observed. Our calculated accretion-powered plateau and fallback timescale ($\sim 3000$\,d) are consistent with the bolometric fading of NGC\,6946-BH1 \cite{Kochanek2023,Beasor2023, suppmat}.

M31-2014-DS1 has more archival data available than NGC\,6946-BH1, which excludes a similar optical outburst and better constrains the fading timescale, which we connect with our hydrogen-depleted progenitor model. Nevertheless, the similarities between M31-2014-DS1 and NGC\,6946-BH1 lead us to conclude they were both associated with the core collapse of massive hydrogen-depleted stars that produced stellar mass BHs. Accretion onto the BHs might produce X-ray emission, but that has not been detected for either M31-2014-DS1 (from archival X-ray observations \cite{suppmat}) and NGC\,6946-BH1, which is explained by X-ray absorption from the surrounding gas \cite{2000ApJ...541..860B, Basinger2021, suppmat}. Theoretical models of BH formation from massive stars predict a wide range of stellar progenitors \cite{Sukhbold2016, Boccioli2024} with few observational constraints. Using previous estimates for the fraction of failed SNe \cite{Adams2017b, Neustadt2021, Byrne2022}, we calculate that the probability of us identifying at least one event in our search is 1 to 20\% \cite{suppmat}. Our calculated stellar evolution of M31-2014-DS1 is similar to many core collapse SN progenitors (Figure \ref{fig:prog}) implying a complex (possibly chaotic) relationship between stellar birth mass and BH formation for stars with initial masses $\gtrsim 12$\,\Msun, as previously predicted on theoretical grounds \cite{Sukhbold2016, Boccioli2024}.

\clearpage



\begin{figure}[!ht]
    \centering
    \includegraphics[width=0.99\textwidth]{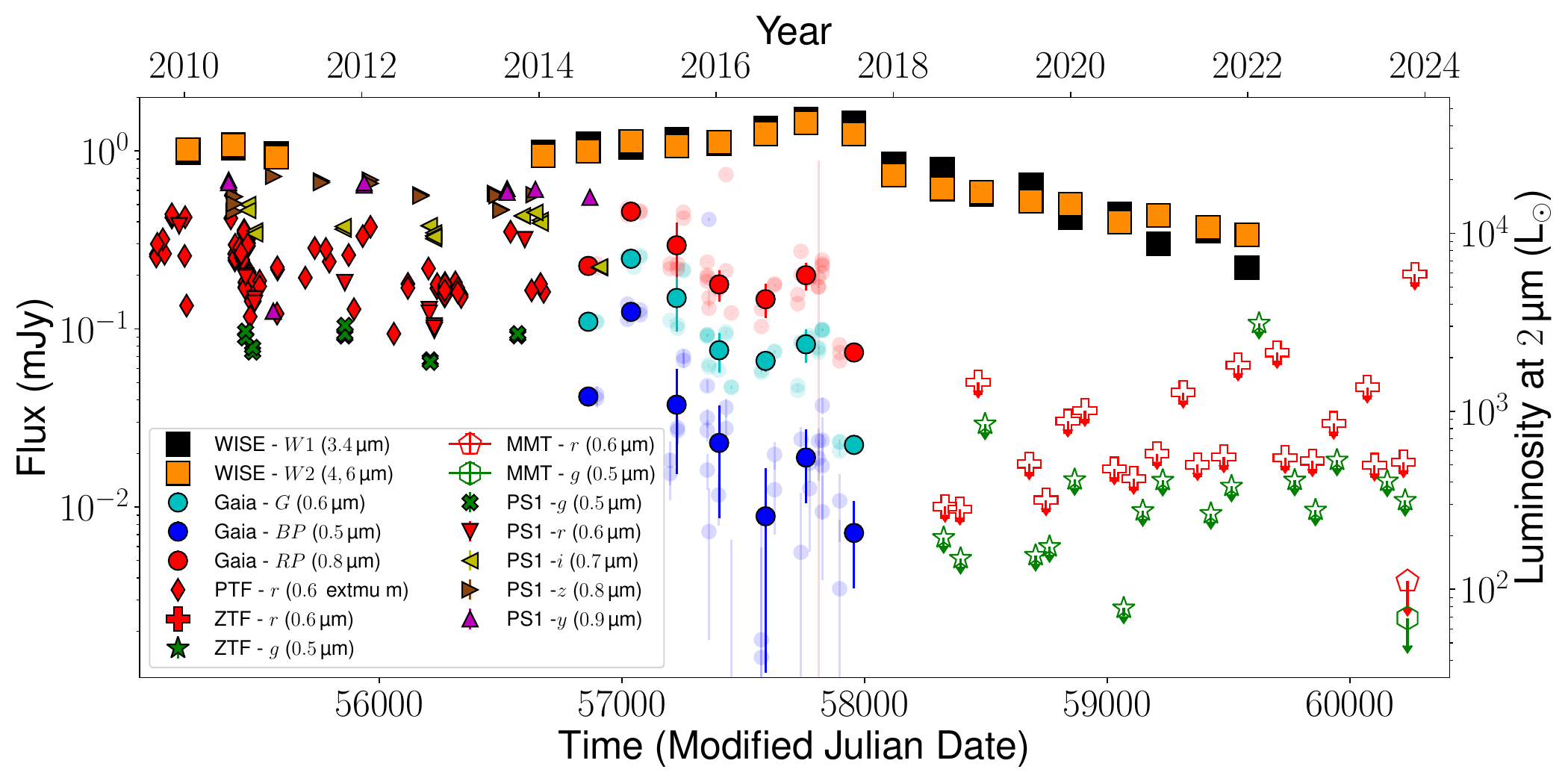}
    \linespread{1.0}\selectfont{}
    \caption{\textbf{Brightness of M31-2014-DS1 as a function of time}. The measured flux in millijansky (mJy, left axis) and equivalent luminosity at the distance of the Andromeda Galaxy (right axis), both on logarithmic scales, are plotted as a function of time in modified Julian date (MJD, lower axis) and Gregorian year (upper axis). Archival data are from the Palomar Transient Factory (PTF), NEOWISE, PS1, Gaia and Zwicky Transient Facility (ZTF) surveys; also shown are follow-up photometric data from the MMT Observatory \cite{suppmat}. Error bars are $1\sigma$ confidence (smaller than the symbol size for the NEOWISE and pre-2014 data); hollow symbols with downwards arrows are $5\sigma$ upper limits. For the Gaia photometry, we show the raw measurements as light symbols, while the dark symbols are the averages within $45$\,d of the closest epoch of NEOWISE data. The luminosity is monochromatic $\lambda\,F_\lambda$, where $\lambda$ is the wavelength and $F_\lambda$ is the flux density, scaled to a wavelength of $2$\,\textmu m. The MIR data are shown on a linear scale in Figure \ref{fig:linearlc}, and all photometric data are provided (with references provided in \cite{suppmat}) in the online data repository.}
    \label{fig:optirlc}
\end{figure}

\begin{figure}
    \centering
    \includegraphics[width=0.49\textwidth]{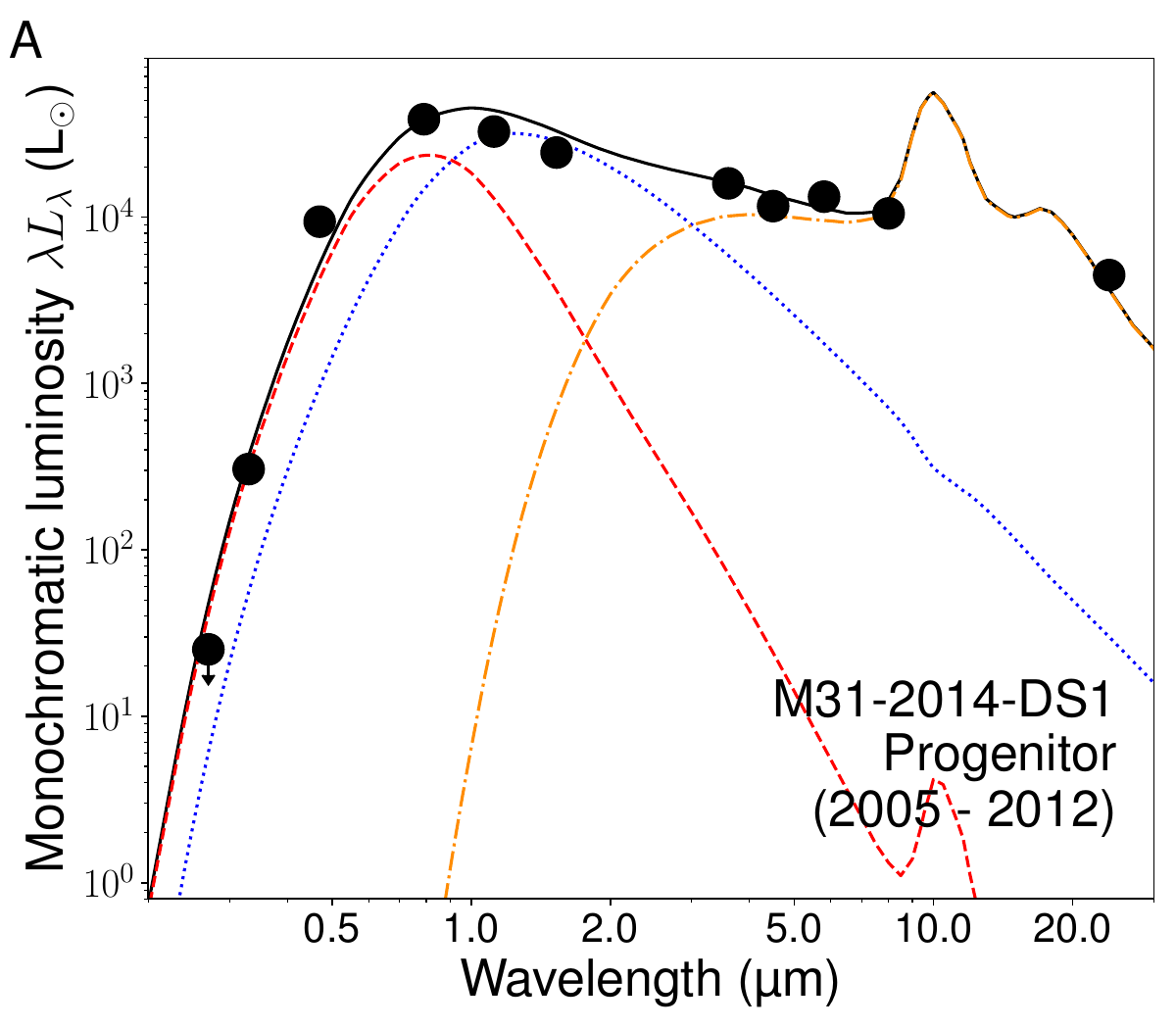}
    \includegraphics[width=0.49\textwidth]{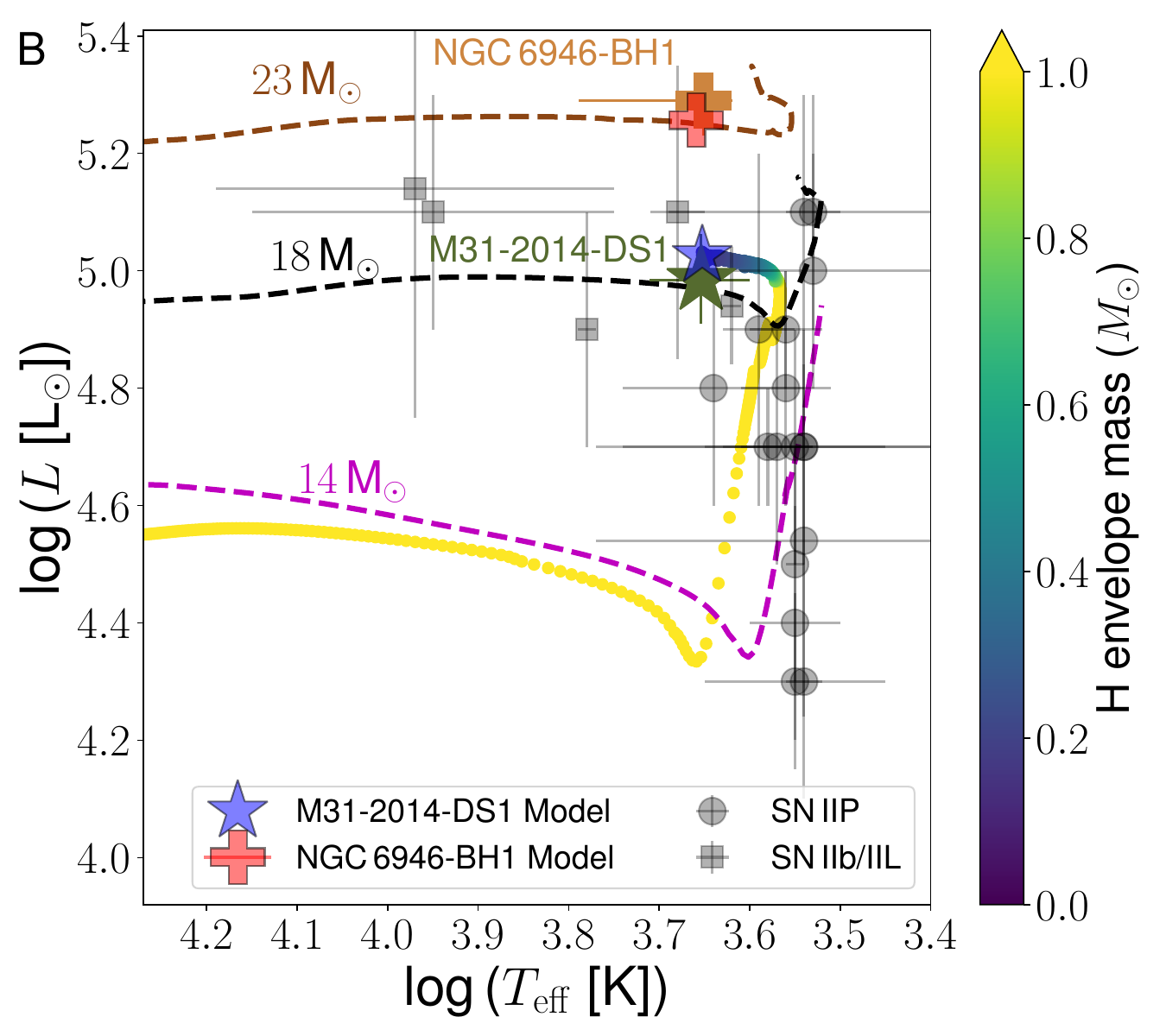}
    \linespread{1.0}\selectfont{}
    \caption{\textbf{Spectral energy distribution and physical properties of the progenitor.} (A) The ultraviolet to MIR SED of M31-2014-DS1 (solid and hollow circles, in units of the monochromatic luminosity $\lambda\,L_{\lambda}$, where $L_\lambda$ is the luminosity density) observed with HST and SST from 2005-2012. Lines are the best-fitting \texttt{DUSTY} model (parameters are listed in Table \ref{tab:dusty_fits}). Total flux (black solid line), dust emission (orange dot-dashed), dust-scattered stellar emission (red-dashed) and dust-attenuated stellar emission (blue dotted). Error bars are $1\sigma$ confidence (smaller than the symbol sizes), and downward arrows are $5\sigma$ upper limits. (B) The observed luminosity and effective temperature of M31-2014-DS1 (green star, from panel A) compared to theoretical single star evolutionary tracks at different initial masses (colored dashed lines) \cite{Choi2016}. The observed progenitor of NGC\,6946-BH1 (brown plus symbol) and those of known hydrogen-rich SNe \cite{Smartt2015} (type indicated in legend) are shown. Error bars are $1\sigma$ confidence. Colored circles are our stellar evolution model for the progenitor of M31-2014-DS1 \cite{suppmat}, with the residual hydrogen envelope mass indicated by the color bar. The model at the time of core collapse is indicated for M31-2014-DS1 (blue star) and for NGC\,6946-BH1 (red plus symbol).}
    \label{fig:prog}
\end{figure}

\begin{figure}
    \centering
    \includegraphics[width=0.49\linewidth]{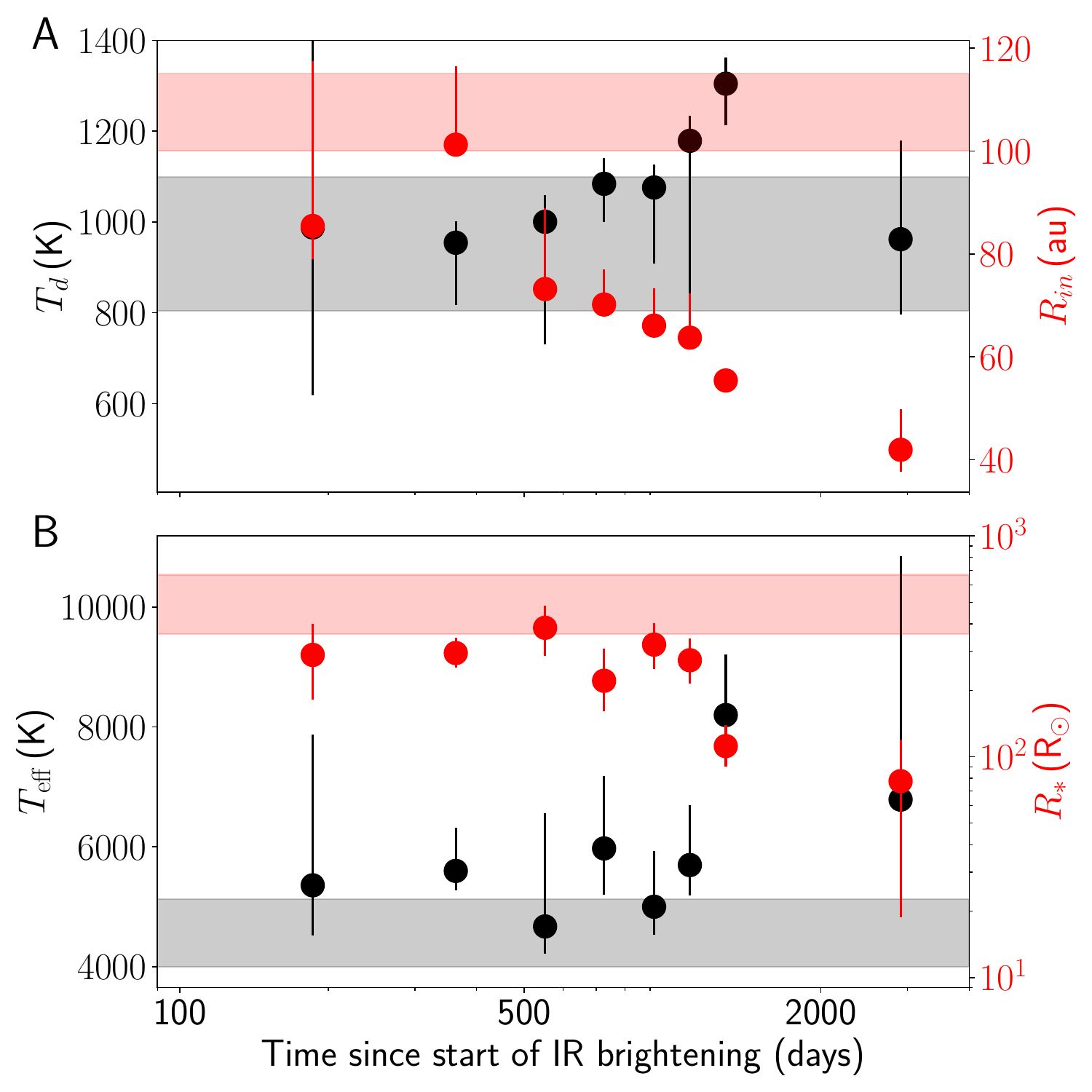}
    \includegraphics[width=0.49\textwidth]{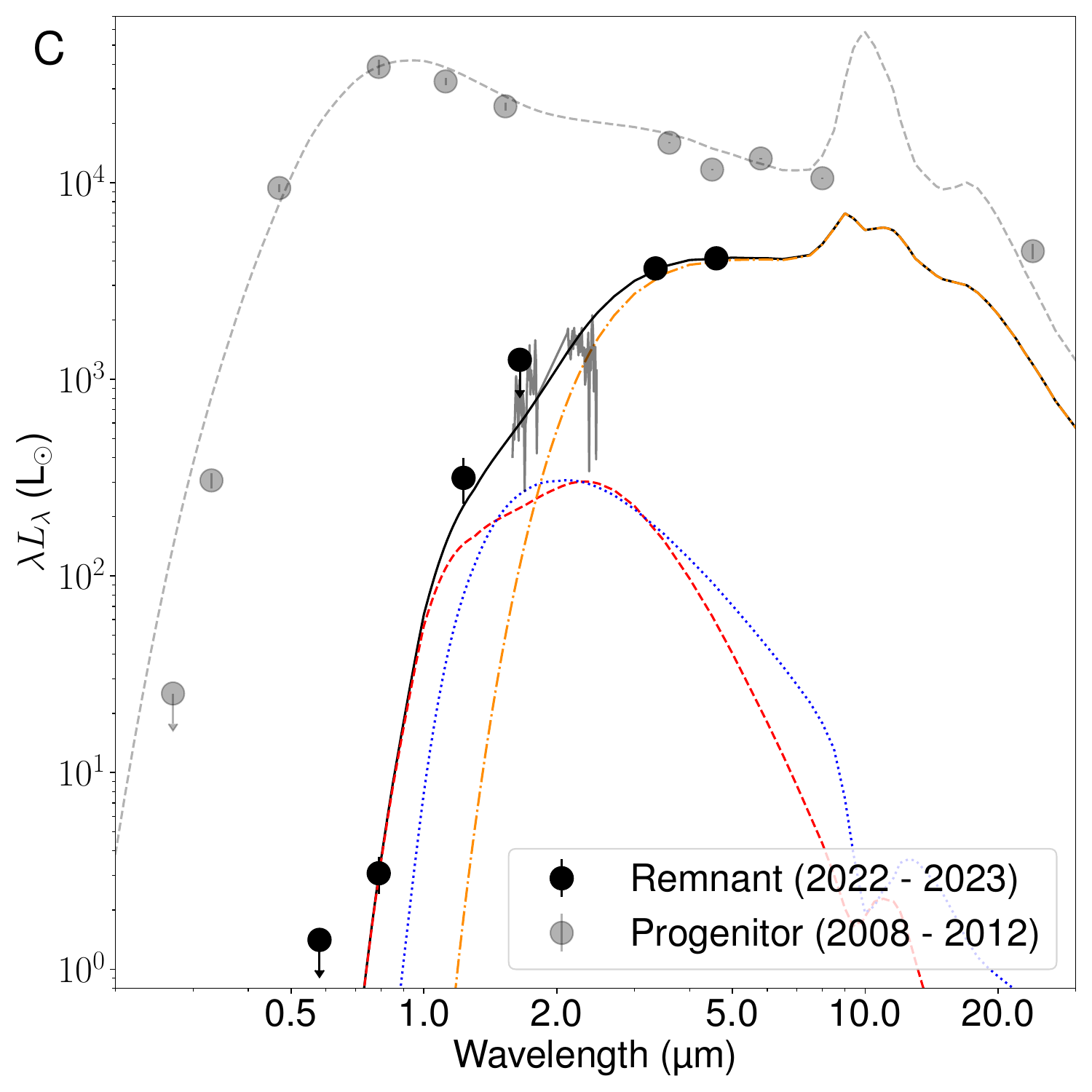}
    \linespread{1.0}\selectfont{}
    \caption{\textbf{Evolution of the stellar and dust properties of M31-2014-DS1.} Temporal evolution of the dust shell and stellar photosphere inferred from the SED model fitting: (A) the dust temperature ($T_d$; black, left axis) and inner shell radius ($R_in$; red, right axis) and (B) the effective temperature (black, left axis) and inferred stellar radius (red, right axis). These are plotted as functions of time since the start of the MIR brightening in 2014 (MJD 56674.19; Figure \ref{fig:optirlc} and \cite{suppmat}). The shaded regions are the corresponding model parameters for the progenitor from 2005-2012. Other parameters are shown in Figure \ref{fig:dust_param_evol}. The SED of the remnant of M31-2014-DS1 (C) in $2022$ to $2023$ [photometry in black circles and spectrum in gray lines (shown on a linear scale in Figure \ref{fig:nires})]. The best-fitting \texttt{DUSTY} model with line styles as in Figure \ref{fig:prog}, and parameters are listed in Table \ref{tab:dusty_fits}. The corresponding progenitor photometry (gray circles) and SED model (dashed grey line) are shown. Error bars are $1\sigma$ confidence and downward arrows are $5\sigma$ upper limits.}
    \label{fig:evol_remnant}
\end{figure}

\begin{figure}
    \centering
    \includegraphics[width=0.49\textwidth]{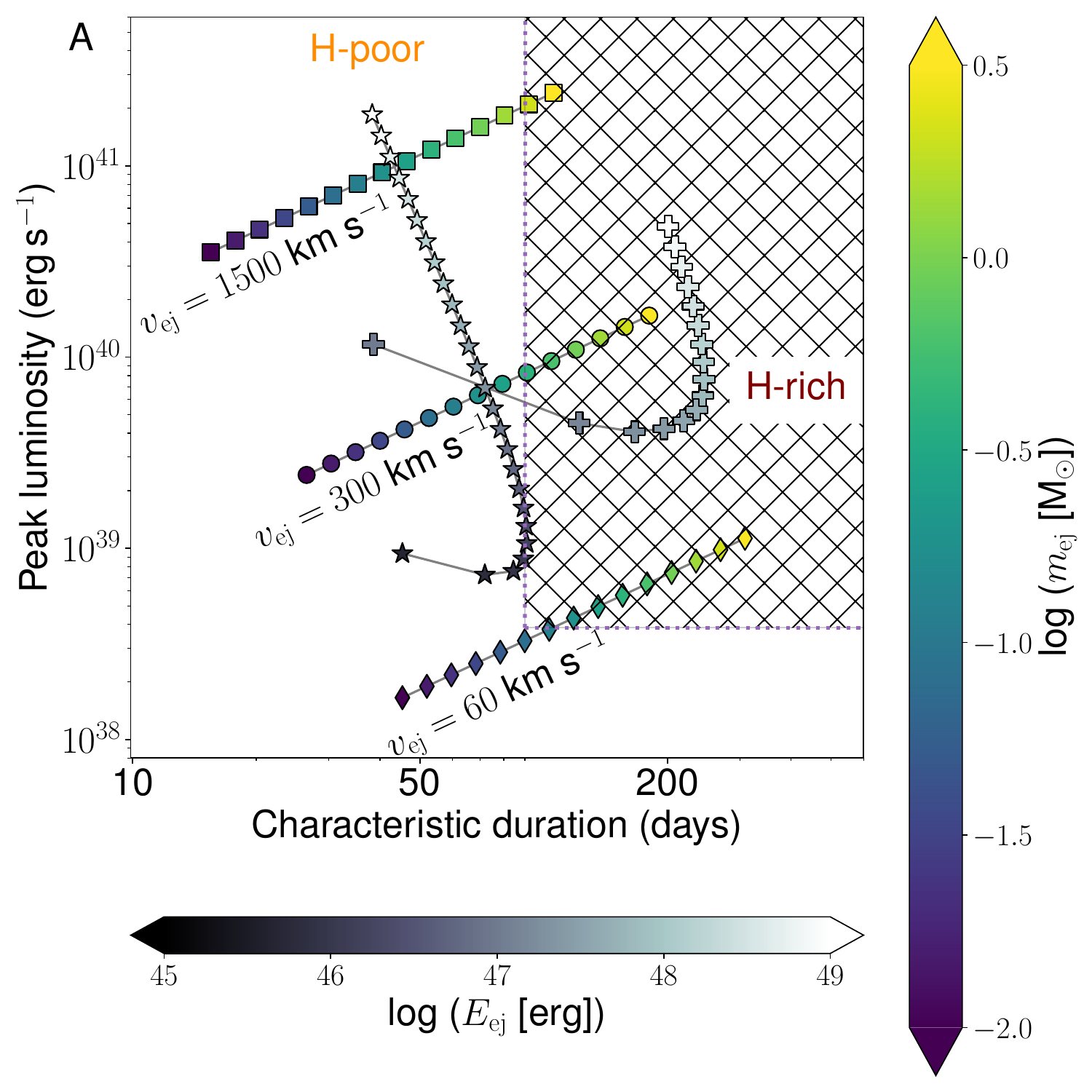}
    \includegraphics[width=0.49\textwidth]{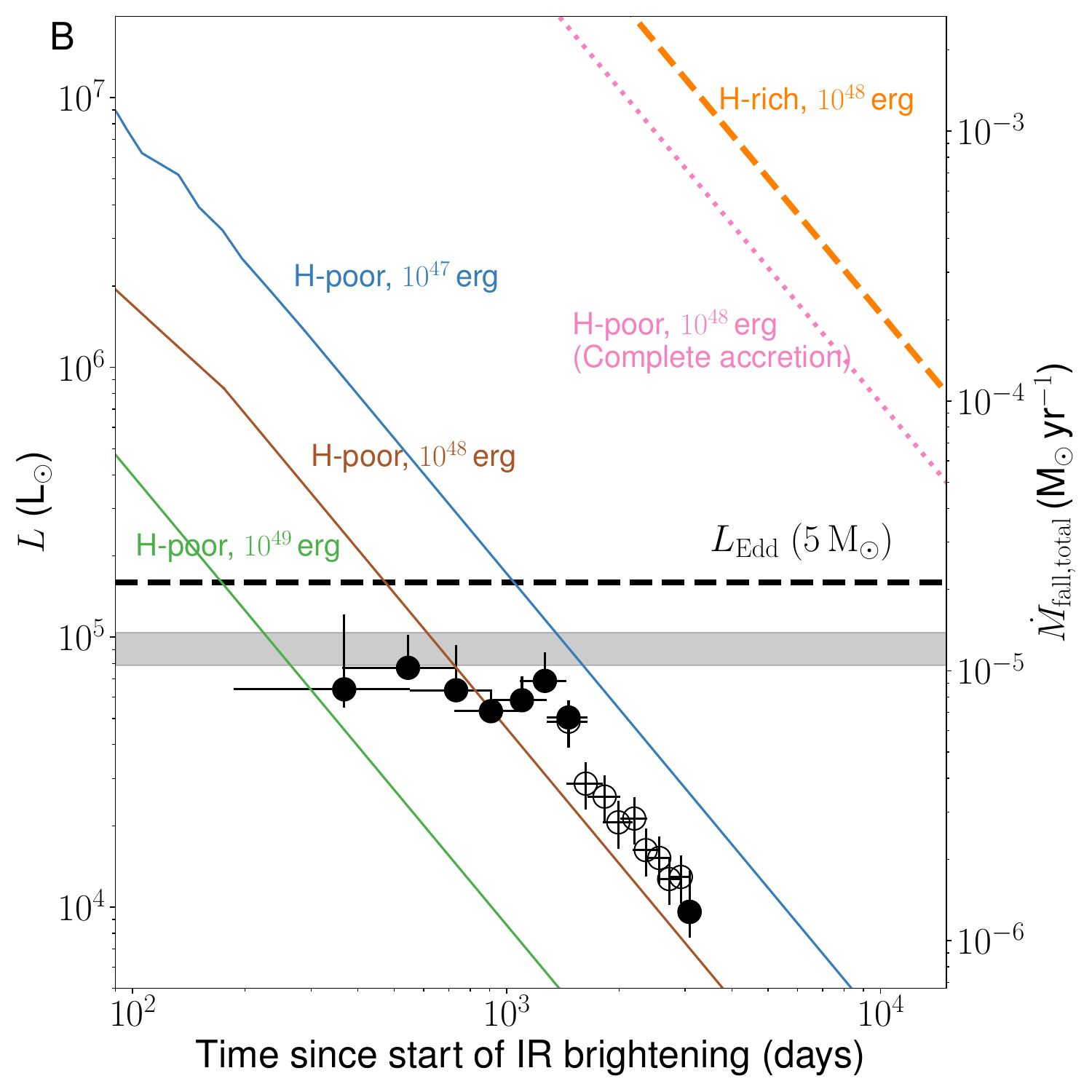}
    \linespread{1.0}\selectfont{}
    \caption{\textbf{Constraints on the mass ejection and stellar envelope fallback in M31-2014-DS1.} (A) The luminosity and duration of transients powered by energy injection into hydrogen-rich envelopes, for ejecta with different velocities ($v_{\rm ej}$) and kinetic energies ($E_{\rm ej}$). The diamonds, circles, and squares indicate ejecta with velocities of $\approx 1\times$, $\approx 5\times$, and $\approx 25\times$ the progenitor's escape velocity respectively, with colors indicating ejecta mass ($m_{\rm ej}$; right color bar). The cross-hatched area shows the phase space ruled out by the optical photometry \cite{suppmat}. Also shown are the luminosity and transient duration for the H-poor (stars) and H-rich (plus symbols) models as a function of ejecta kinetic energy (bottom color bar). (B) The bolometric fading of M31-2014-DS1 compared to models of mass fallback rates for different explosion energies. The hydrogen-poor and hydrogen-rich progenitor models both assume $5$\% accretion radiative efficiency. For comparison, the dotted pink line shows $100$\% accretion in the H-poor case and the dashed orange line shows an example H-rich case. Black solid circles are parameters inferred from the SED models, and hollow circles are estimates using a bolometric correction to MIR luminosity \cite{suppmat}. Error bars are $1\sigma$ confidence. The gray shaded region shows the progenitor luminosity, and the black dashed line shows the Eddington luminosity ($L_{\rm Edd}$) for a $5$\,\Msun BH. The right axis shows the corresponding total mass fallback rate ($\dot{M}_{\rm fall, total}$) from the models.}
    \label{fig:model}
\end{figure}


\clearpage 

%
\bibliography{science_template} 
\bibliographystyle{sciencemag}

%
%
%
%
%
%


\section*{Acknowledgments}
We thank the anonymous referees for a careful review of the manuscript and providing constructive feedback for improving its contents. We thank H. Gupta, K. Das and P. Nair for assistance with the observations. We thank S. R. Kulkarni, L. Hillenbrand, J. Goldberg and C. Conroy for valuable discussions. Data were obtained at the W. M. Keck Observatory from telescope time allocated to the National Aeronautics and Space Administration through the agency's scientific partnership with the California Institute of Technology and the University of California. The Observatory was made possible by the generous financial support of the W. M. Keck Foundation. The authors wish to recognize and acknowledge the very important cultural role and reverence that the summit of Maunakea has always had within the indigenous Hawaiian community. We are most fortunate to have the opportunity to conduct observations from this mountain. This research has made use of the Keck Observatory Archive (KOA), which is operated by the W. M. Keck Observatory and the NASA Exoplanet Science Institute (NExScI), under contract with the National Aeronautics and Space Administration. We acknowledge the Visiting Astronomer Facility at the Infrared Telescope Facility, which is operated by the University of Hawaii under contract 80HQTR19D0030 with the National Aeronautics and Space Administration.
\paragraph*{Funding:}
K. D. was supported by NASA through the NASA Hubble Fellowship grant \#HST-HF2-51477.001 awarded by the Space Telescope Science Institute, which is operated by the Association of Universities for Research in Astronomy, Inc., for NASA, under contract NAS5-26555. K. D.'s data analysis was supported by NASA through a Keck PI Data Award, administered by the NASA Exoplanet Science Institute and through ADAP grant number 80NSSC24K0663. M.MacLeod was supported by a Clay postdoctoral fellowship at the Smithsonian Astrophysical Observatory. A.A. was supported by the Simons Foundation through a Flatiron Research Fellowship. E.Q.'s work benefited from interactions at workshops funded by the Gordon and Betty Moore Foundation through grant GBMF5076. 

\paragraph*{Author contributions:}
K.D. identified the object, initiated follow-up observations, carried out the analysis and wrote the manuscript. M.MacLeod developed theoretical models and helped write the manuscript. J.E.J. analyzed the archival space telescope data and R.M.L. assisted with the observational interpretation. A.A., A.L., E.L. and E.Q. contributed to the theoretical interpretation and modeling. M.M.K. and R.S. assisted with ground-based observational follow-up. E.K., M.Masterson and C.P. performed the X-ray data analysis. A.M.M. produced the time-resolved stacks of WISE images. All authors contributed to the scientific interpretation.

\paragraph*{Competing interests:}
There are no competing interests to declare.

\paragraph*{Data and materials availability:}

The photometric data, image subtraction tools and theoretical stellar evolution models and failed supernova calculations are available at  \url{https://github.com/dekishalay/M31-2014-DS1/} and archived on Zenodo \cite{zenodo}. The NEOWISE images can be accessed at \url{http://byw.tools/wiseview} by providing the source coordinates. The image subtraction photometry code used for the NEOWISE images can be found at \url{https://github.com/dekishalay/M31-2014-DS1/} and Zenodo \cite{zenodo}. The Gaia photometry can be accessed at \url{https://vizier.cds.unistra.fr/viz-bin/VizieR-3?-source=I/355/epphot} by providing the source coordinates. The PTF light curves can be accessed at \url{https://irsa.ipac.caltech.edu/cgi-bin/Gator/nph-scan?submit=Select&projshort=PTF} by providing the source coordinates. The ZTF light curves can be produced at \url{https://ztfweb.ipac.caltech.edu/cgi-bin/requestForcedPhotometry.cgi} by providing the source coordinates. The MMT images can be accessed at \url{https://www.mmto.org/cfa-optical-infrared-science-archive/} using the source coordinates. The PanSTARRS light curves can be accessed at \url{https://catalogs.mast.stsci.edu/panstarrs} by providing the source coordinates. The HST images can be accessed at \url{https://mast.stsci.edu/portal/Mashup/Clients/Mast/Portal.html} by providing the source coordinates and using Program IDs 12111, 14072 and 16730. The Spitzer photometry can be accessed at \url{https://vizier.cds.unistra.fr/viz-bin/VizieR?-source=II/368} by providing the source coordinates. The Keck images and spectra can be accessed at \url{https://koa.ipac.caltech.edu/cgi-bin/KOA/nph-KOAlogin} using Program IDs 2023B\_N258 and 2023B\_N258. The IRTF data can be accessed at \url{https://irsa.ipac.caltech.edu/applications/irtf/} using Program ID 2023B053.


\subsection*{Supplementary materials}
Materials and Methods\\
Supplementary Text\\
Figs. S1 to S12\\
Tables S1 to S2\\
References \textit{(59-\arabic{enumiv})}\\ 


\newpage


\renewcommand{\thefigure}{S\arabic{figure}}
\renewcommand{\thetable}{S\arabic{table}}
\renewcommand{\theequation}{S\arabic{equation}}
\renewcommand{\thepage}{S\arabic{page}}
\setcounter{figure}{0}
\setcounter{table}{0}
\setcounter{equation}{0}
\setcounter{page}{1} 


\begin{center}
\section*{Supplementary Materials for\\ \scititle}

Kishalay De$^\ast$, Morgan MacLeod, Jacob E. Jencson, Elizabeth Lovegrove,\and
Andrea Antoni, Erin Kara, Mansi M. Kasliwal, Ryan M. Lau, Abraham Loeb,\and
Megan Masterson, Aaron M. Meisner, Christos Panagiotou, Eliot Quataert, \and and Robert Simcoe\and

\small$^\ast$Corresponding author. Email: kd3038@columbia.edu
\end{center}

\subsubsection*{This PDF file includes:}
Materials and Methods\\
Supplementary Text\\
Figs. S1 to S12\\
Tables S1 to S2\\
References \textit{(59-\arabic{enumiv})}

\newpage


\subsection*{Materials and Methods}

\subsubsection*{Identification in NEOWISE}

The Wide-field Infrared Survey Explorer (WISE) satellite\cite{Wright2010}, and its extended mission NEOWISE\cite{Mainzer2014}, performed all-sky MIR survey in the $W1$ ($3.4$\,\textmu m) and $W2$ ($4.6$\,\textmu m) bands since 2014; we use data up to 2022. NEOWISE revisits each part of the sky once every $\approx 0.5$\,yr. We searched for transients in time-resolved coadded images produced as part of the unWISE project \cite{Lang2014,Meisner2018}. We used the ZOGY algorithm\cite{Zackay2016} to perform image subtraction on the NEOWISE images using the co-added images of the WISE mission (obtained in 2010-2011) as reference images. The pipeline produces a database of all transients with statistical significance of $\gtrsim 10\sigma$. Follow-up for the sources was coordinated using the \texttt{fritz} astronomical data platform\cite{vanderWalt2019}. 

We applied selection criteria to the catalog of WISE transients to identify luminous MIR outbursts spatially coincident with M31 and M33 satisfying the following criteria:
\begin{enumerate}
    \item Transient located within 4 degrees of the nucleus of M31 or M33.
    \item The source brightened from the reference image and had no prior history of outbursts or variability in WISE data before the flagged brightening. The transient was detected in at least two consecutive epochs, to reject foreground moving objects.
    \item The transient was not coincident with a foreground Galactic star, determined from the parallax and proper motion measurements from Gaia Data Release 3\cite{GaiaEDR3}. 
    \item The transient has peak MIR luminosity of $\gtrsim 10^{37}$\,erg\,s$^{-1}$, chosen to select outbursts that reach the expected Eddington luminosity for a $\approx 0.1$ to $1$\,M$_\odot$ star, while excluding lower luminosity IR outbursts from young stars and X-ray binaries.
\end{enumerate}

\begin{figure}[!t]
    \centering
    \includegraphics[width=0.99\linewidth]{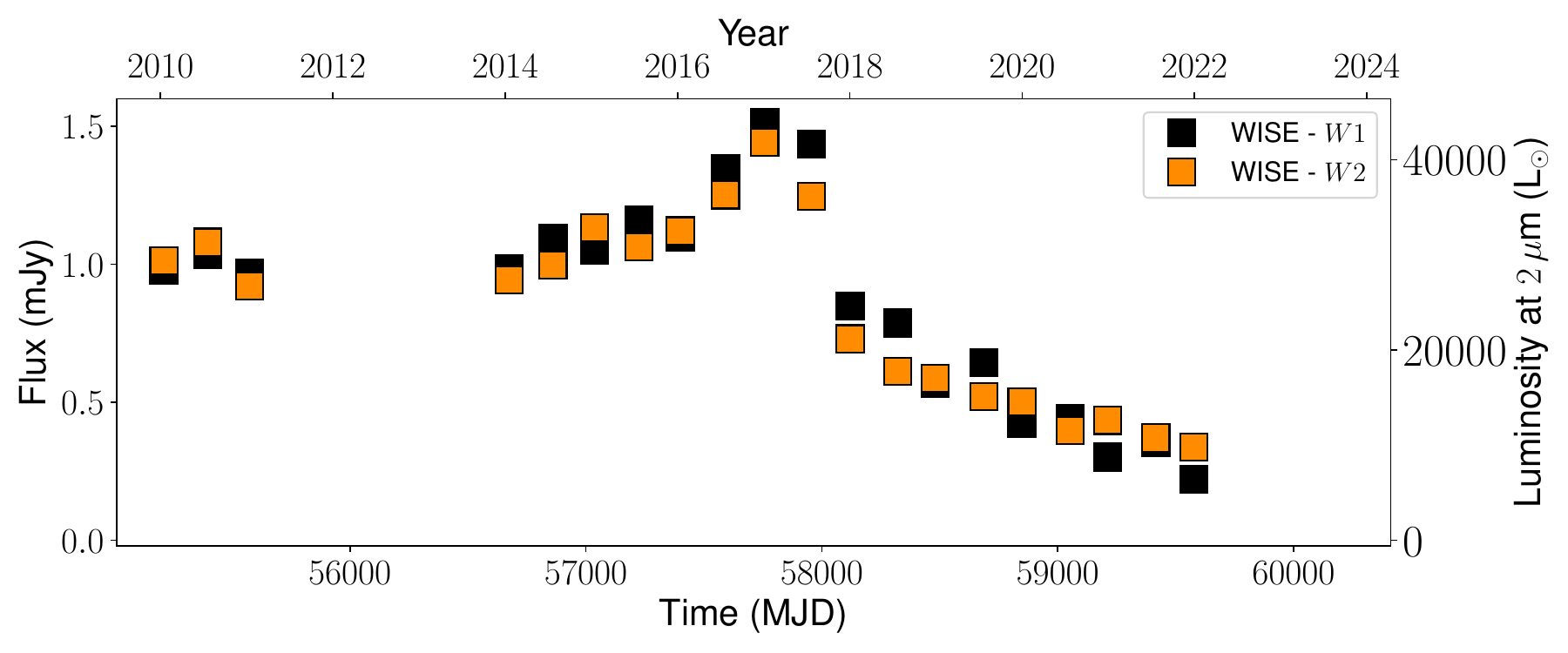}
    \caption{{\bf NEOWISE light curve of M31-2014-DS1}. Same as Figure \ref{fig:optirlc}, but showing only the NEOWISE data and on a linear vertical scale.}
    \label{fig:linearlc}
\end{figure}

We identified the source WTP\,16aathhy (internal alphanumeric code in the database), initially as a candidate nova, at a projected separation of $\approx 0.54^\circ$ ($\approx 7$\,kpc) from the nucleus of M31. The transient was identified at $> 10\sigma$ significance in difference imaging for the NEOWISE observation of the field in 2016 July (MJD 57591.79) at a (host subtracted) magnitude of $W2 = 17.91 \pm 0.10$\,AB\,mag. The transient is coincident with an optical point source in archival ground-based imaging and Gaia\,DR3. Given the lower spatial resolution of WISE, we adopt the Gaia coordinates of the optical source. The optical source  parallax is consistent with zero ($= 0.09 \pm 0.23$\,mas) as are the proper motions in the right ascension ($\mu_{\rm RA} = -0.17 \pm 0.17$) and declination ($\mu_{\rm Dec} = -0.10 \pm 0.21$) \cite{GaiaEDR3}, indicating it is not a foreground Galactic star. We performed forced photometry on the NEOWISE difference images at the Gaia position, finding the IR brightening of the source began in the first NEOWISE visit of the field on MJD 56674.19 (Figure \ref{fig:linearlc}), followed by a brightening by 50\% over the next $\approx 1000$\,d.  We nominally adopt the time of the first NEOWISE visit as the reference time for our analysis, and designate the source as M31-2014-DS1 based on the onset year of the event and its subsequent identification as a disappearing star. The exact time of the start of the IR brightening is not well constrained.

\subsubsection*{Archival photometry}

The progenitor of M31-2014-DS1 has been reported in historical archival optical and infrared source catalogs. It was previously considered as an irregular variable star in the field of M31\cite{Kaluzny1998} and designated as V7984\,M31B. We searched for archival time resolved photometry of the source from ground and space-based surveys. The field was observed with the PTF\cite{Law2009, Rau2009} between 2010 and 2014 in the $r$-filter; we retrieved time resolved photometry of the source from the survey archive\cite{Ofek2012}. The source was observed in the $grizy$ filters by the PS1 survey \cite{Chambers2016} between 2010 and 2014; we retrieved the point source photometry from the online source catalog\cite{Flewelling2020}. We retrieved optical photometry of the source between 2014 and 2017 in the Gaia $G$, $BP$ and $RP$ passbands from the Gaia-Andromeda survey\cite{Evans2023} in Gaia DR3. This was observed as part of the Zwicky Transient Facility (ZTF\cite{Bellm2019a}) optical survey between 2018 and 2024, but there is no source detected at that position in $g$ and $r$-bands, in either the stacked reference image or individual epochs of observation. We derive $5\sigma$ upper limits on the source flux from forced Point Spread Function (PSF) photometry on the ZTF difference images\cite{Masci2019}, and stacking the non-detection upper limits in bins of 60\,d. The archival light curve is shown in Figure \ref{fig:optirlc}, and all photometry is provided in the online data repository.

The source was serendipitously observed by the HST and SST at multiple epochs between 2005 and 2022. We analyzed the available HST imaging from 2012 taken with the Advanced Camera for Surveys (ACS; F475W, F814W filters) and Wide Field Camera 3 (WFC3) in both the Ultraviolet-Visible (UVIS; F336W, F275W filters) and Infrared (IR; F110W, F160W filters) channels as part of the Panochromatic Hubble Andromeda Treasury (PHAT) program\cite{Dalcanton2012}, as well as the 2016 ACS/Wide Field Camera (WFC) imaging in F625W and 2022 WFC3/UVIS imaging in F606W and F814W. We processed the available HST images with \texttt{DOLPHOT}\cite{Dolphin2000,Dolphin2016} to obtain PSF-fitting photometry of M31-2014-DS1. As inputs to \texttt{DOLPHOT}, we use the Charge Tranfer Efficiency-corrected \texttt{flc} frames for ACS/WFC and WFC3/UVIS and \texttt{flt} frames for WFC3/IR downloaded from the Mikulski Archive for Space Telescopes. We adopt the \texttt{DOLPHOT} parameter settings for each camera as used by PHAT\cite{Dalcanton2012,Williams2014}. We ran \texttt{DOLPHOT} separately for each instrument and filter combination over the available epochs, using the drizzled 2012 F814W mosaic as a common reference image for alignment across all runs. \texttt{DOLPHOT} computes and applies aperture corrections to a radius of $0.5$\arcsec\ for the reported photometry. We then corrected to infinite apertures for each instrument and filter combination \cite{Bohlin2016,STScI_WFC3_PhotCal_2025}. We retrieved SST photometry of the source in 2005 from the Spitzer source list\cite{Spitzer2021}. The photometry was acquired with the Infrared Array Camera (IRAC \cite{Fazio2004}) in Channels 1 ($3.6$\,\textmu m), 2 ($4.5$\,\textmu m), 3 ($5.8$\,\textmu m) and 4 ($8.0$\,\textmu m), and using the Multiband Imaging Photometer for Spitzer (MIPS \cite{Rieke2004}) in Channel 1 ($24.0$\,\textmu m). 
We use the measurements from 2012 and 2022 to characterize the progenitor and remnant of M31-2014-DS1, and list them in Table \ref{tab:prog_photo}.

\subsubsection*{Follow-up imaging}

We obtained follow-up optical imaging of the source using the Binospec instrument\cite{Fabricant2019} on the MMT telescope. Observations were obtained between Coordinated Universal Time (UTC) 2023 October 18 and 2023 October 19, consisting of dithered exposures of the field with total exposure time of 2400\,s and 1800\,s in the $g$ and $r$ filters respectively. The data were reduced, stacked and calibrated against the PS1 catalog. We performed aperture photometry at the position of the source, and derive $5\sigma$ upper limits of $r> 22.44$\,mag and $g > 22.96$\,mag, limited primarily by source confusion in the dense field.

We obtained near-infrared imaging of the source using the SpeX instrument\cite{Rayner2003} on the NASA Infrared Telescope Facility on UTC 2023 September 06. A series of dithered exposures were obtained with a total exposure time of 1200\,s in both $J$ and $K$ filters. The images were reduced, stacked and calibrated against the 2MASS catalog. No source is detected at the transient position, with a $5\sigma$ limiting magnitude of $J > 19.0$\,mag and $K> 17.8$\,mag.

We obtained deeper near-infrared imaging using the Multi-Object Spectrometer For Infrared Exploration (MOSFIRE\cite{McLean2012}) on the Keck-I telescope, on UTC 2023 December 25. The source was observed over multiple dithered exposures with a total exposure time of 288\,s and 297\,s in $J$ and $K$ bands respectively. High winds during the $K$-band observations led us to discard those images. Performing aperture photometry at the source position, a faint source is detected in the $J$-band images at $J = 20.90 \pm 0.28$\,Vega\,mag. 

\subsubsection*{Follow-up infrared spectroscopy}

\begin{figure}
    \centering
    \includegraphics[width=0.99\linewidth]{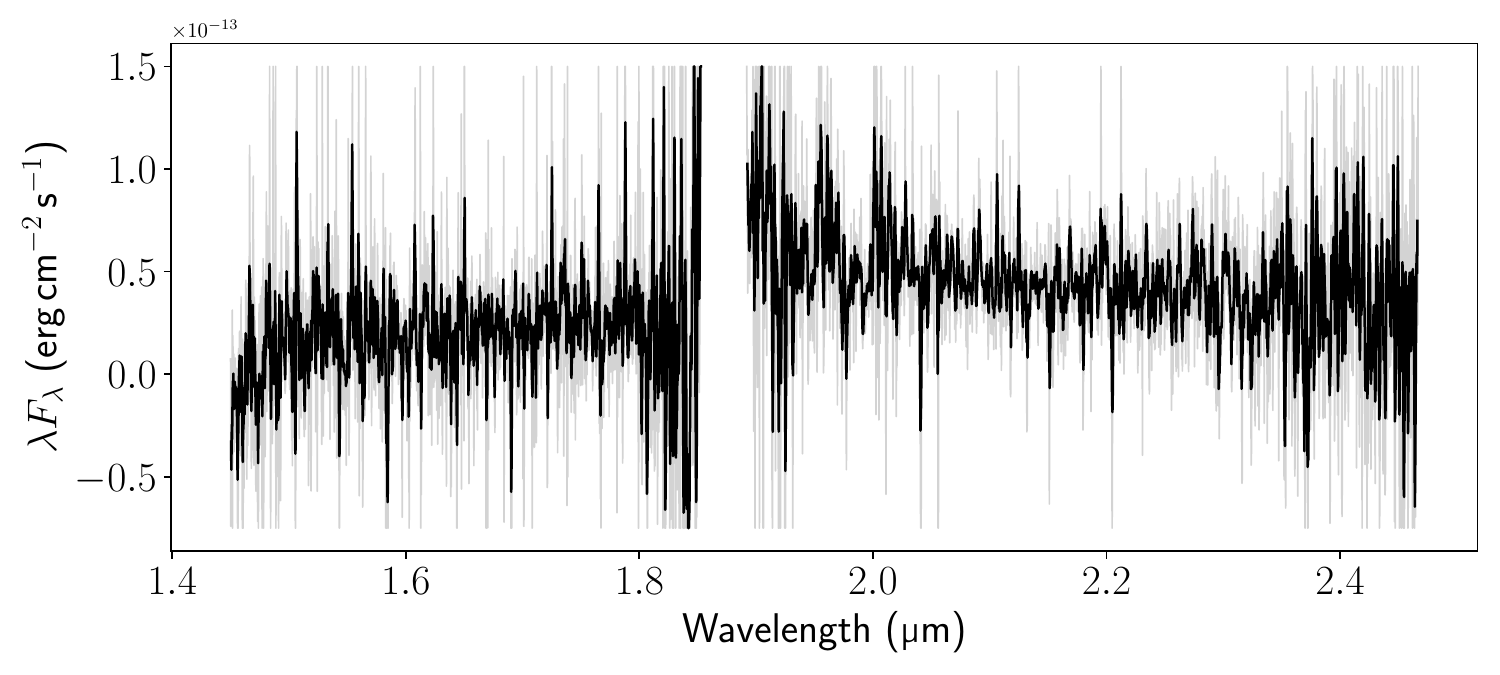}
    \caption{{\bf NIRES spectrum of M31-2014-DS1}. Grey lines show the raw spectrum and black lines show the spectrum binned by averaging every five points to improve the signal-to-noise ratio.}
    \label{fig:nires}
\end{figure}

On UTC 2023 September 04, we obtained follow-up $1.0 - 2.5$\,\,\textmu m near-infrared spectroscopy of the source using the Near Infrared Echelle Spectrometer (NIRES\cite{Wilson2004}) on the Keck-II telescope. Due to the faintness of the source, it was acquired by centering a nearby bright star in the slit followed by applying a blind offset using the Gaia coordinates. We obtained a series of dithered exposures in the ABBA pattern with a total exposure time of 2400\,s. We reduced the data using the \texttt{spextool}\cite{Cushing2004} software, followed by flux calibration and telluric correction using the \texttt{xtellcor}\cite{Vacca2003} package. A faint source is detected in the spectral trace in $H$ and $K$ bands, with a nearly featureless spectrum brightening towards longer wavelengths, shown in Figure \ref{fig:evol_remnant} and Figure \ref{fig:nires}.

\subsubsection*{X-ray observations}

To constrain the X-ray luminosity of M31-2014-DS1, we retrieved archival observations of the source location by the Chandra X-ray telescope. The source was within the field of view of a 50~ks {\it Chandra} Advanced CCD Imaging Spectrometer-I observation in 2015. We reduced the data using \textsc{Ciao}\cite{Fruscione2006} (version 4.16) and CALDBv4.11.0 \cite{CXCCALDB}, finding no source at the Gaia position. Using the \texttt{aplimits} tool \cite{Fruscione2006} and following published procedures, we estimate a 3$\sigma$ upper limit on the X-ray flux $F_X \lesssim 2.0 \times 10^{-15}$~erg~s$^{-1}$~cm$^{-2}$ in the 0.5-7 keV band. This limit was calculated using a circular source region with a radius of 5\arcsec\ and by assuming a power-law spectrum (number of photons $N \propto E^{-\Gamma}$, where $E$ is the photon energy and $\Gamma = 2$ is the assumed photon index) with Galactic absorption\cite{HI4PICollaboration2016} corresponding to a hydrogen column density $N_H = 2.26 \times 10^{21}$~cm$^{-2}$. At a distance of 770 kpc \cite{Savino2022}, this corresponds to an X-ray luminosity $L_X \lesssim 1.4 \times 10^{35}$~erg~s$^{-1}$.

We also searched the Swift archive for observations within 12\arcmin\ of the source after 2015, finding four observations in July 2020. Using the XRT products generator \cite{Evans2007,Evans2009}, we stacked these observations ($\approx 4$~ks in total) and set a 3$\sigma$ upper limit of $F_X \lesssim 1.3 \times 10^{-13}$~erg~s$^{-1}$~cm$^{-2}$ in the 0.3-10 keV band, assuming the same power-law as for the Chandra observations. Without correcting for any intrinsic absorption, this corresponds to $L_X \lesssim 8.9 \times 10^{36}$~erg~s$^{-1}$.

The Gaia position was also observed serendipitously by NuSTAR four times in 2015. We followed standard procedure \cite{Madsen2015} to reduce the observational data using the \textit{nupipeline} and \textit{nuproducts} tools of the \textit{NuSTAR} Data Analysis Software (NuSTARDAS) package version 2.1.2 \cite{Madsen2015} and the corresponding CALDBv20240422 \cite{Madsen2022}. No source at the Gaia position is detected in any of the observations. Assuming the same power-law spectrum, we estimated a stacked 3$\sigma$ upper limit 2-10 keV flux $F_X \lesssim 2.2 \times 10^{-14}$~erg~s$^{-1}$~cm$^{-2}$, which corresponds to $L_X \lesssim 1.6 \times 10^{36}$~erg~s$^{-1}$.

\subsubsection*{\texttt{DUSTY} modeling}

M31-2014-DS1 exhibits excess MIR emission throughout its evolution from 2014 to 2022, indicating dust surrounding the star. We model the optical to MIR SED of the transient to estimate the properties of the central star and circumstellar dust shell, and their evolution over time. We use the dust radiative dust transfer code \texttt{DUSTY} V4\cite{Ivezic1997, Ivezic1999} to produce models that are fitted to the multi-wavelength data. We assume a spherically symmetric distribution of the dust with a radial density profile that changes with radius $r$ as $\propto r^{-2}$ around the star, which is assumed to be a point source. We assume the dust grains are composed of warm silicates, as expected for massive supergiants \cite{Verhoelst2009}, and a grain size distribution from \cite{Mathis1977} -- with minimum and maximum grain sizes of $a_{\rm min} = 0.005$\,\textmu m and $a_{\rm max} = 0.25$\,\textmu m.

\begin{figure}[!t]
    \centering
    \includegraphics[width=\textwidth]{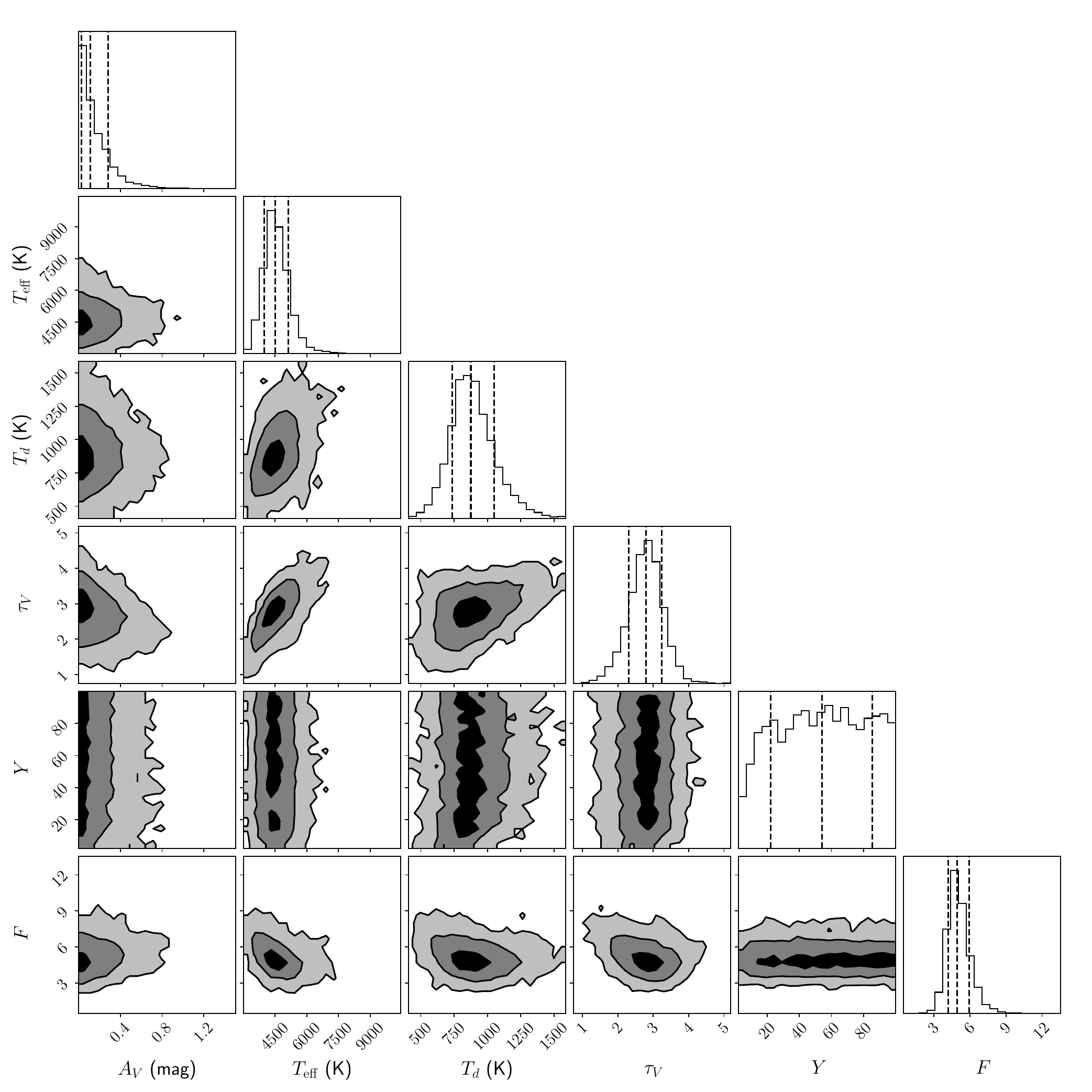}
    \linespread{1.0}\selectfont{}
    \caption{{\bf Corner plot for the MCMC DUSTY modeling of the progenitor of M31-2014-DS1 in 2012.} $F$ is in units of $10^{-12}$\,erg\,cm$^{-2}$\,s$^{-1}$. The dark, medium and light shaded regions indicate contours enclosing 39\%, 86\% and 99\% of the probability region for the respective parameters. The median of the posterior distributions and the corresponding 68\% confidence intervals are shown as vertical dashed lines in the one dimensional histograms and listed in Table \ref{tab:dusty_fits}.}
    \label{fig:prog_dusty}
\end{figure}

\begin{figure}[!t]
    \centering
    \includegraphics[width=\textwidth]{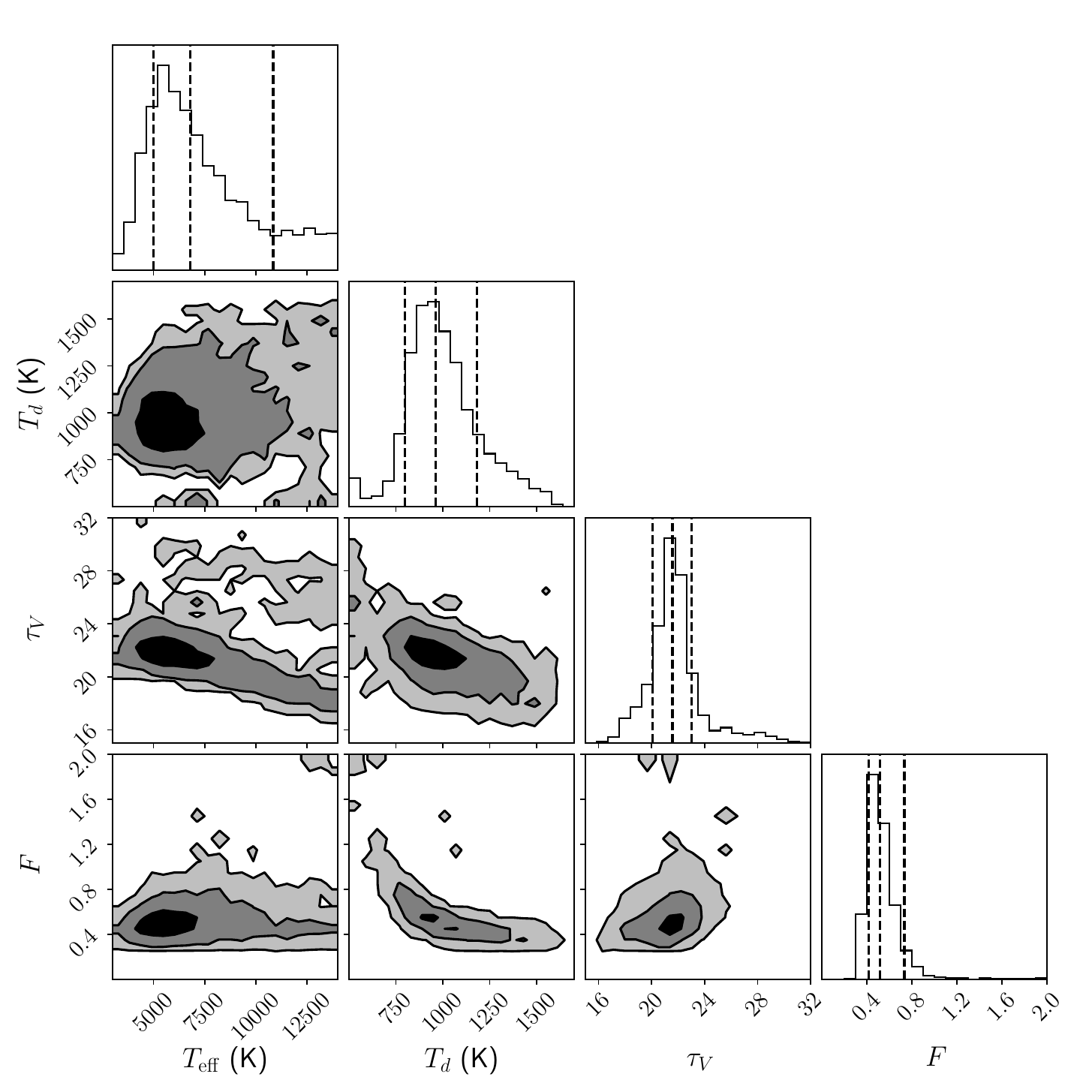}
    \linespread{1.0}\selectfont{}
    \caption{{\bf Corner plots for the MCMC DUSTY modeling of the remnant of M31-2014-DS1 in 2022-2023.} Same as in Figure \ref{fig:prog_dusty} but for the SED in 2022-2023. The median parameter values and their 68\% confidence intervals are listed in Table \ref{tab:dusty_fits}; $A_V$ and $Y$ were fixed to their 2012 values.}
    \label{fig:rem_dusty}
\end{figure}

We fitted the progenitor photometry from HST and SST using a Markov Chain Monte Carlo (MCMC) wrapper around the \texttt{DUSTY} code\cite{De2022} using the Python \texttt{emcee} library\cite{Foreman-Mackey2013}. We model the foreground wavelength dependent interstellar extinction using a standard relation \cite{Fitzpatrick1999} extending from ultraviolet to the MIR. The free parameters of the model are the dust optical depth at $0.55$\,\,\textmu m ($\tau_V$), the foreground visual extinction ($A_V$), the inner stellar temperature ($T_*$), the dust temperature at the inner edge of the shell ($T_d$), the ratio of the outer to inner radius of the shell ($Y$) and the total flux ($F$). We assume flat priors on all the free parameters and ensure convergence of the posterior sampling chains by checking for stabilization of the autocorrelation time to $< 1$\% in consecutive steps. The posterior probability distributions are shown in Figure \ref{fig:prog_dusty} and listed in Table \ref{tab:dusty_fits}.

We find the thickness $Y$ to be poorly constrained because the wavelength coverage is primarily confined to the shorter wavelength MIR bands where the emission is dominated by the hotter, inner part of the dust shell. The best-fitting foreground extinction is consistent with the expected Milky Way extinction in this direction ($A_V \approx 0.1$\,mag)\cite{Schlafly2011}, with no additional extinction within M31. To constrain the SED of the remnant, we perform the same analysis using the HST, Keck and NEOWISE data obtained during $\approx 2022 - 2023$, fixing the foreground extinction to $A_V = 0.1$\,mag and the thickness $Y = 10$ based on the progenitor model. Consistent with the optical disappearance and slow infrared fading, we find the remnant has a much higher optical depth and $\approx 10\times$ fainter bolometric flux (Figure \ref{fig:rem_dusty} and Table \ref{tab:dusty_fits}). 

\begin{figure}[!t]
    \centering
    \includegraphics[width=\textwidth]{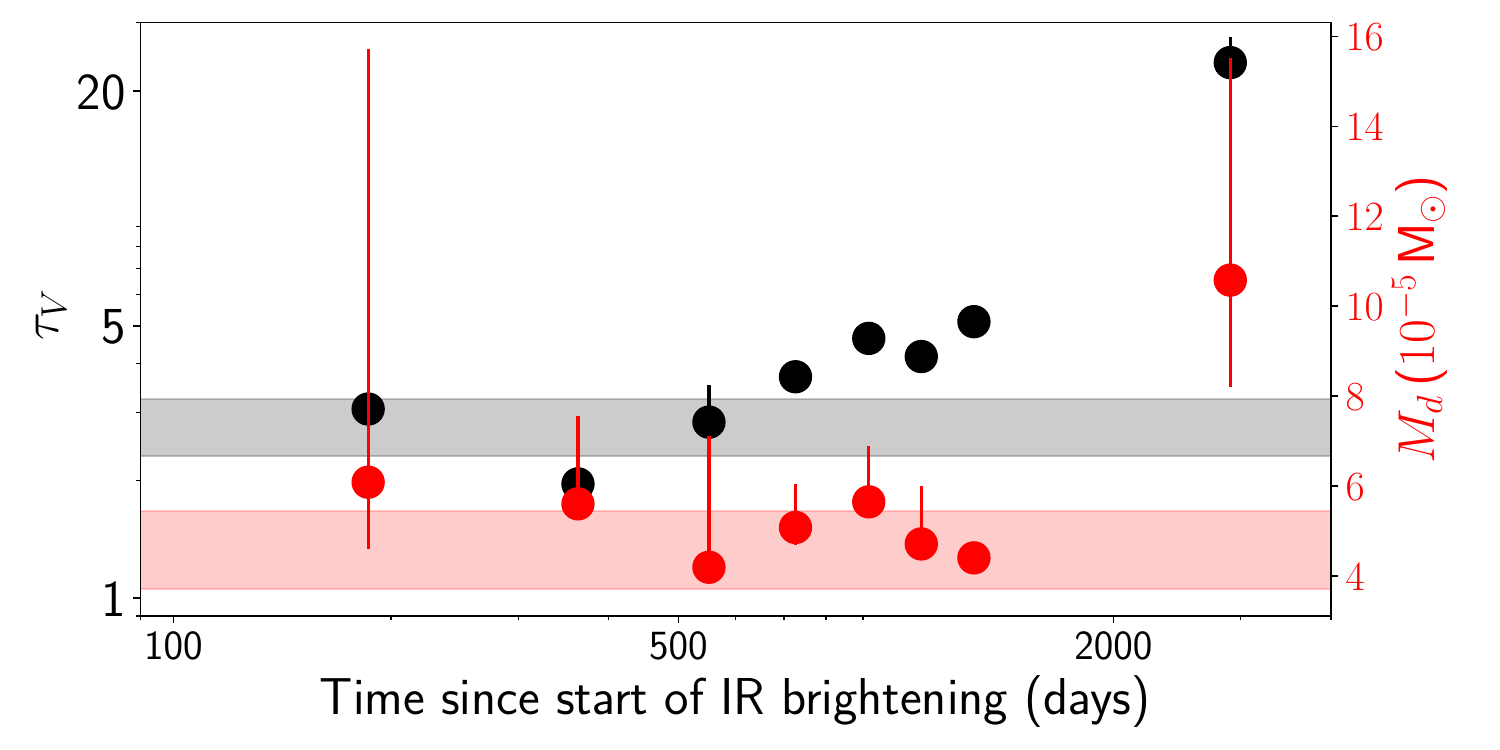}
    \linespread{1.0}\selectfont{}
    \caption{{\bf Optical depth and total mass of circumstellar dust during the dimming episode}. Errors are shown at $1\sigma$ confidence level. Same as Figure \ref{fig:evol_remnant}A, but for the optical depth (black, left axis) and mass of dust in the circumstellar shell (red, right axis).}
    \label{fig:dust_param_evol}
\end{figure}

 A stellar companion to the massive progenitor star \cite{Sana2012} could affect both the pre-disappearance and remnant photometry. Any companion is unlikely to be a similarly evolved cool star, given the very short lifetime of this phase \cite{Neugent2020}. For unevolved companions, lower mass (smaller than A-type dwarfs; $\lesssim 2.5$\,\Msun) stars are unlikely because the total lifetime of the RSG would be shorter than the pre-main sequence lifetime of the companion \cite{Neugent2018}, while compact stellar remnants (e.g. a companion neutron star/BH) would have no observable effects on the observed photometry. Therefore, OB-type stellar companions are the only possible contaminants for the observed photometry, which would contribute excess emission at the bluest wavelengths ($\lesssim 5000$\,\AA; \cite{Neugent2018}) because the cool supergiant outshines the companion in redder light \cite{Neugent2021}. Previous analysis of stellar model grids has shown that such excess blue emission is identifiable in color-color space of $U-B$ and $R-I$ \cite{Neugent2018}.

Based on the photometry of the progenitor \cite{Massey2016}, we find $U-B = 1.30 \pm 0.01$ and $R-I = 0.72 \pm 0.01$. The $U-B$ color is much redder than selection thresholds for RSG + OB type binaries ($U-B < 1$; \cite{Neugent2020}) and inconsistent with the entire binary RSG photometry model grid \cite{Neugent2018}. Those models did not include yellow supergiants (YSGs) or RSGs with dust shells, but those would produce even bluer $U-B$ colors due to the hotter YSG star or decreased contribution of RSG emission at bluer wavelengths, respectively. Therefore, we find no evidence for a binary companion in the progenitor photometry could affect the SED. A-type and lower mass stars are not likely given evolutionary age constraints; because stars in this lower mass range have optical luminosity $\lesssim 10$\,\Lsun, the observed optical fading of the star to already a few\,\Lsun (Figure \ref{fig:evol_remnant}) disfavors the effects of a binary companion on the remnant SED as observed in 2022-2023. 

\subsubsection*{Dust evolution during the disappearance}

To investigate the evolution of the source, we use the Gaia multi-color photometric coverage during the optical fading (Figure \ref{fig:optirlc}) and the NEOWISE photometry. Because the Gaia measurements are not simultaneous with the NEOWISE coverage, we average the Gaia measurements around the time of each NEOWISE observation (within 1.5 months of each epoch; Figure \ref{fig:optirlc}). For each NEOWISE epoch, we construct an optical (Gaia $BP$ and $RP$) to MIR (NEOWISE $W1$ and $W2$) SED. We then perform the same SED fitting for these epochs as for the remnant modeling above. At each epoch, we estimate the total mass of dust in the shell as
\begin{equation}
    M_d = \frac{4 \pi \tau_V R_{\rm in}^2 Y}{\kappa_d} r_{\rm dg}
\end{equation}
where $R_{\rm in}$ is the radius of the inner shell, $\kappa_d \approx 50$\,cm$^2$\,g$^{-1}$ is the average visual opacity per unit total mass (gas and dust) and $r_{\rm dg} \approx 0.01$ is the assumed dust-to-gas mass ratio relevant for the late-phase evolution of low energy explosions \cite{Karambelkar2025, Kaminski2021}. Adopting $Y=10$, we find
\begin{equation}
    M_d \approx 6.1 \times 10^{-8} {\rm M}_\odot \left(\frac{R_{in}}{1000\,{\rm R}_\odot}\right)^2 \tau_V 
\end{equation}
The inferred progenitor optical depth indicates a wind-loading parameter $A = \dot{M}/4\pi\,v_w \approx 7 \times 10^{13}$\,g\,cm$^{-1}$, where $\dot{M}$ and $v_w$ are the progenitor mass-loss rate and wind velocity, respectively. Taking $v_w \approx 50 - 100$\,km\,s$^{-1}$, typical for YSGs with our inferred $T_{\rm eff}$ ($\approx 4500$\,K; \cite{Humphreys2023}), implies a high mass-loss rate of $\approx (1-2) \times 10^{-4}$\,\Msun\,yr$^{-1}$.
The subsequent evolution of the dust temperature, inner radius, inner source temperature and inferred photospheric radius are shown in Figure \ref{fig:evol_remnant}, while the evolution of the dust optical depth and total mass are shown in Figure \ref{fig:dust_param_evol}. The evolution of the total source luminosity during this time is shown in Figure \ref{fig:model}.

To estimate the source luminosity evolution between the end of the Gaia coverage in 2017 and the SED of the remnant in 2022, we estimate a bolometric correction between the MIR luminosity using trapezoidal integration of the NEOWISE fluxes and the total luminosity from the \texttt{DUSTY} models for the last Gaia + NEOWISE epoch in 2017. We apply this bolometric correction to the NEOWISE MIR data from 2018 to 2022, which is shown in Figure \ref{fig:model}. 

This analysis shows a gradual increase in the dust temperature and a decrease in the inner shell radius (Figure \ref{fig:evol_remnant}A) and total luminosity, from the progenitor to the remnant in 2022. A similar trend is seen in the evolution of the stellar temperature and radius, with the remnant having a higher inferred temperature and smaller radius of $\approx 20$\% of the progenitor star. We compare the evolution of the dust shell radius to the minimum radius at which dust can form around a luminous star, $r_c$, given by\cite{Kochanek2011}
\begin{equation}
    r_c \approx 11\,{\rm au} \left(\frac{L}{10^5\,{\rm L}_\odot}\right)^{1/2} \left(\frac{1500\,{\rm K}}{T_d}\right)^2 Q_{rat}^{-1/2}
\end{equation}
where $L$ is the source luminosity and $Q_{rat}$ is the ratio of average absorption efficiency of the dust to the stellar photosphere weighted by the blackbody spectrum of the star. For $T_d \approx 1500$\,K and $T_* \approx 5000$\,K, the analytical approximation in \cite{Kochanek2011} provides $Q_{rat} \approx 0.15$, so
\begin{equation}
    r_c \approx 28\,{\rm au} \left(\frac{L}{10^5\,{\rm L}_\odot}\right)^{1/2} \left(\frac{1500\,{\rm K}}{T_d}\right)^2
\end{equation}
We attribute the inferred contraction of the inner shell radius to $\lesssim 50$\,au (Figure \ref{fig:evol_remnant}A) to dust formation near the condensation radius, causing the apparent shell to shrink in radius starting from the progenitor circumstellar shell. The optical depth of the dust shell increases by a factor of $\gtrsim 10\times$ due to the gradual contraction of the shell, but the total dust mass remains roughly constant in the initial $\approx 1000$\,d before increasing by $\approx 10^{-4}$\,\Msun during the subsequent period as the source evolves to the 2022 remnant.

\subsubsection*{Constraints on mass ejection}

\begin{figure}[!t]
    \centering
    \includegraphics[width=\linewidth]{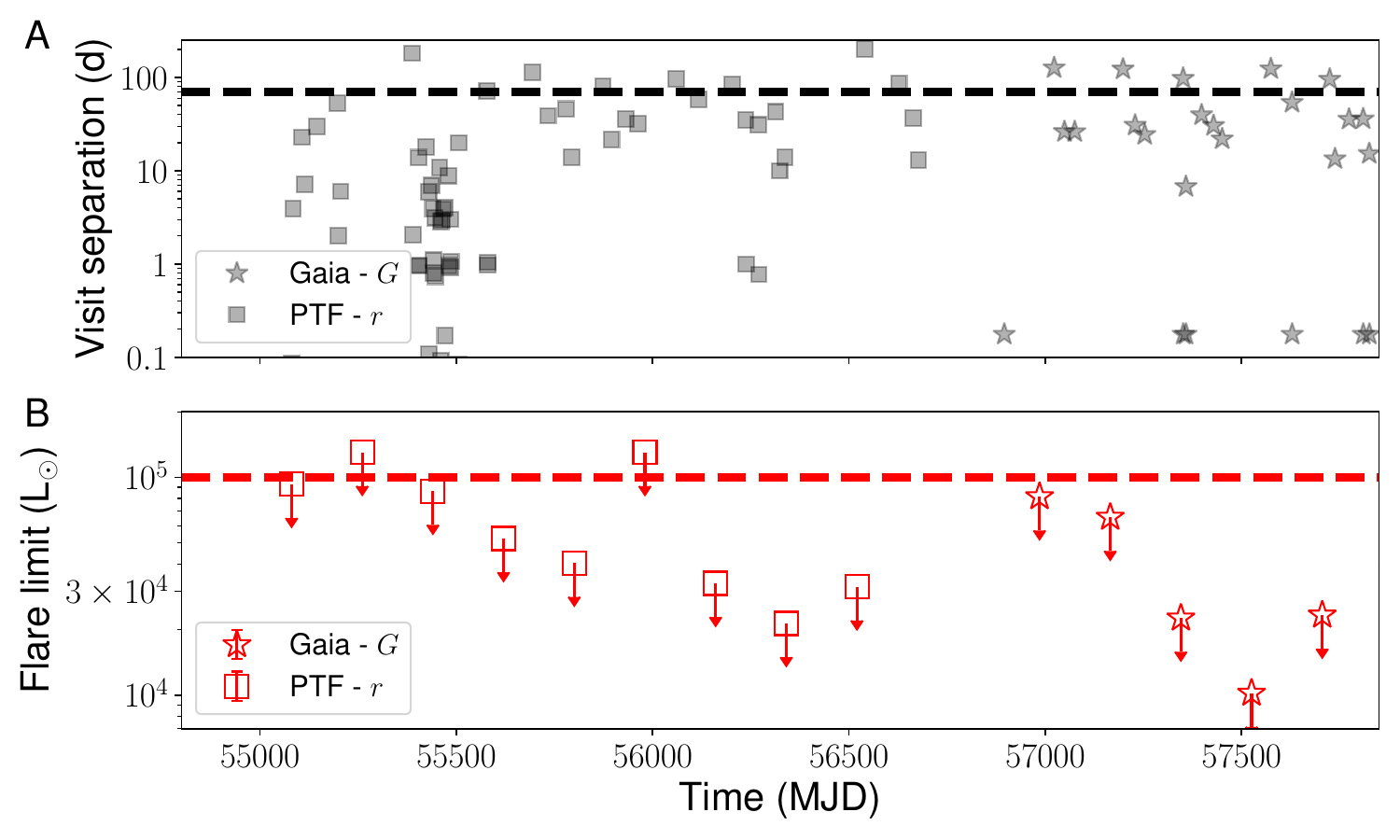}
    \linespread{1.0}\selectfont{}
    \caption{{\bf Constraints on the peak luminosity and timescale of an optical outburst in M31-2014-DS1.} (A) Time separation between successive observations of the source in PTF $r$-band (squares) and Gaia (stars) observations. (B) Downward arrows are $10\sigma$ upper limits on the luminosity of an optical outburst after accounting for the irregular variability of the star (see text). The dashed lines show the adopted upper limits on the timescale (panel A) and luminosity (panel B) of a missed outburst.}
    \label{fig:cadence_lim}
\end{figure}

We use the continuous optical photometric coverage of M31-2014-DS1 between 2010 and 2020 by PTF and Gaia to place limits on any optical outburst, which is predicted for H-recombination powered transients in failed SNe\cite{Lovegrove2013}. The progenitor light curve (Figure \ref{fig:optirlc}) does not show an obvious optical outburst. To quantify the timescale of a possibly missed outburst, we measure the difference between successive observations of the star in PTF $r$-band and Gaia $G$-band data, which provide the most frequent temporal sampling. We use the visit separations over this period to quantify the timescale of an outburst that could have been missed between successive observations (Figure \ref{fig:cadence_lim}). We adopt an upper limit on the rise timescale of any missed outburst as half of the maximum visit separation during this time period $\approx 180$\,d. As the source exhibits irregular variability, which could hide a low amplitude outburst, we measure the standard deviation of the flux variability over six-month bins , and adopt an upper limit on the luminosity of a missed outburst as $10\sigma$ deviation from the median luminosity of the source in these bins (Figure \ref{fig:cadence_lim}). We adopt half of the highest luminosity limit during this period ($\approx 10^5$\,\Lsun) as an upper limit on the luminosity of any missed optical outburst in M31-2014-DS1. These constraints on the outburst luminosity and timescale are shown in Figure \ref{fig:model}A as the hatched area.

Stellar mass shock-heated and expelled in a full or partial stellar explosion expands, cools, and radiates as it becomes transparent, producing transient but luminous emission. In the case of hydrogen-rich ejecta like that of Type II SNe, stellar merger ejecta, or weak explosions of massive stars, emission is dominated by the recombination of hydrogen at temperatures of a few to ten thousand kelvin. Empirical scaling relations characterize the timescale, color, and luminosity of such H-rich ejecta, though depend upon the stellar model and shock passage through the star\cite{2010MNRAS.405.2113D,2019ApJ...879....3G}. In the case of failed SNe, because the initial infall, shock passage, and subsequent ejection of material happen over one to several stellar dynamical times, we expect transients that are broader but less luminous than their fully impulsive counterparts\cite{Antoni2023}. We adopt the scaling relations [\cite{Kleiser2014}, their equations 13 and 14]:
\begin{equation}
     L_{\rm pl} \approx 1.2\times10^{42} {\rm erg~s}^{-1} \times E_{51}^{5/6}  M_{10}^{-1/2}  R_{500}^{2/3}  \kappa_{0.4}^{-1/3}  T_{6000}^{4/3},
\end{equation}
where $L_{\rm pl}$ is the transient's plateau luminosity, and  
\begin{equation}
    t_{\rm pl} \approx  120~d \times E_{51}^{-1/6}  M_{10}^{1/2}  R_{500}^{1/6}  \kappa_{0.4}^{1/6}  T_{6000}^{-2/3},
\end{equation}
where $t_{\rm pl}$ is the plateau duration. $E_{51}$ is the outburst energy in units of $10^{51}$~erg, $M_{10}$ is the ejecta mass in units of $10$\,\Msun, $R_{500}$ is the progenitor radius in units of $500$\,\Rsun, $\kappa_{0.4}$ is the characteristic opacity before recombination in units of $0.4$\,cm$^2$\,g$^{-1}$, and $T_{6000}$ is the effective temperature at which the recombining material becomes transparent, in units of $6000$\,K. 

We apply this model to determine the range of transient durations and luminosities in Figure \ref{fig:model}. For a typical YSG with a mass of $5M_\odot$ and radius of $500R_\odot$ \cite{Humphreys2023}, the escape velocity is $\approx 60$~km~s$^{-1}$. We therefore adopt representative velocities for ejecta above these minimum escape speeds for different stellar models. Figure \ref{fig:model} shows tracks of $v_{\rm ej} =  60$, $300$ and $1500$\,km\,s$^{-1}$, with different ejecta masses ranging from $10^{-2}\,M_\odot$ to $3\,M_\odot$. Higher ejecta masses or faster velocities correspond to higher energy events that produce more luminous transients. Figure \ref{fig:model}A shows the parameter space of models that are excluded because they would have been detected, given our constraints on the luminosity and duration of an outburst.

\subsubsection*{Progenitor Stellar Models}

\begin{figure}[!t]
    \centering
    \includegraphics[width=0.49\textwidth]{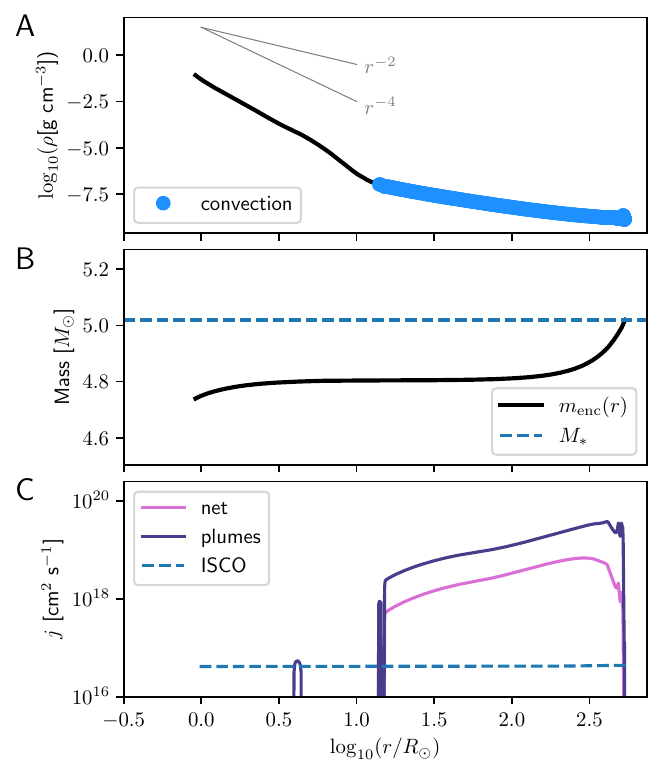}
    \includegraphics[width=0.49\textwidth]{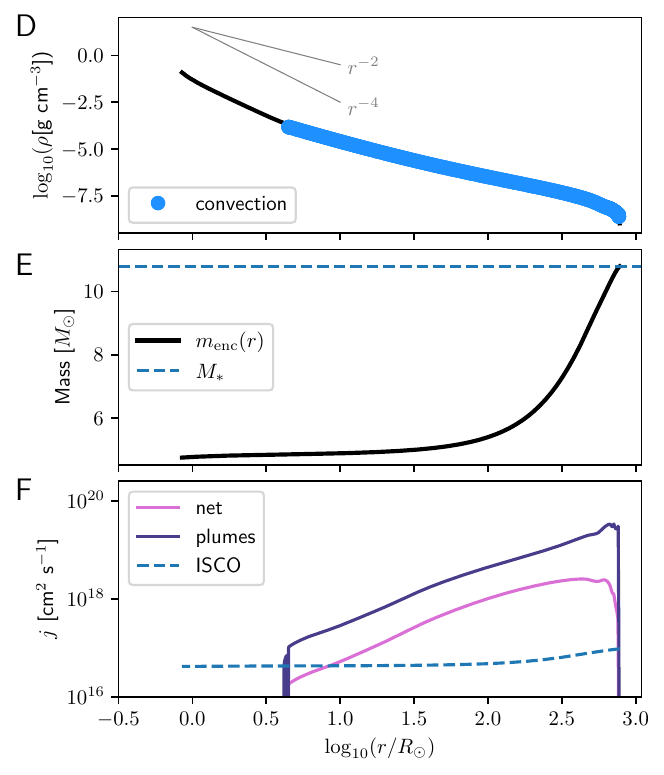}
    \linespread{1.0}\selectfont{}
    \caption{{\bf Comparison of the hydrogen envelope structures of H-poor and H-rich progenitor star models.} (A, D) Radial profile of the mass density in the envelope $\rho$ (black lines), with convective regions highlighted as blue thick regions. Radial density profiles scaling with radius $r$ as $r^{-2}$ and $r^{-4}$ are shown for comparison. (B, E) Radial profiles of the enclosed mass $m_{\rm enc}(r)$ (black lines) with the total model mass shown as a blue horizontal dashed line. (C, F) Radial profiles of the typical angular momentum content $j$ -- the net angular momentum (pink lines) and the characteristic angular momentum of convective plumes (purple lines). The blue dashed line indicates $j$ at the innermost stable circular orbit if all of the enclosed mass were to collapse to a BH. We find that convective regions carry much more random angular momentum than a BH horizon, preventing direct fallback of these outer envelope regions. }
    \label{fig:profiles}
\end{figure}

We model the implosion of a stellar envelope and the resulting accretion onto a central compact object using the stellar evolution code MESA\cite{Paxton2011,Jermyn2023} (version 24.08.1). MESA solves the equations of hydrostatic stellar structure and evolves them forward in time by computing the nuclear and thermal time-evolution of a star. We use these models to investigate about the interior density profile of the progenitor star of M31-2014-DS1. 

The detailed structure of the envelopes of massive stars is debated, especially because most models are limited to one dimension and the hydrostatic approximation \cite{2023Galax..11..105J}. We therefore investigate only the broad differences in stellar structure between stars with large remaining hydrogen envelopes and those largely stripped of hydrogen. By adjusting the model mass loss history, we produce two qualitatively different models that share the approximate luminosity of the observed progenitor. Both models are evolved from the pre-main sequence until carbon core depletion. After carbon core depletion, we expect additional burning stages to continue in the core, but with little effect on the envelope structure.

\paragraph{\it Hydrogen-poor model:} 
For the hydrogen-poor yellow supergiant model, we remove the majority of a partially-evolved star's hydrogen envelope. Physically, this might happen through phases of enhanced wind or binary mass stripping and exchange. The properties of a model star depend on how and when mass loss occurs. There is no evidence of a binary companion for M31-2014-DS1, but there is evidence for a high mass loss rate of the progenitor $\sim 10^{-4}M_\odot$~yr$^{-1}$, required to produce the circumstellar material that reddens the progenitor spectral energy distribution (Figure 3). We therefore consider a MESA model with an elevated mass loss rate in its late evolution as follows.

We begin with a pre-main-sequence stellar model with initial mass $13\,M_\odot$, which is evolved through subsequent burning stages. We adopt a heavy element mass fraction $Z=0.014$, which is approximately solar. Until core helium exhaustion, we apply a nominal stellar mass loss rate using MESA's `Dutch' wind scheme, appropriate for massive stars \cite{Renzo2017}, with a coefficient of unity. The properties of convection affect our modeling and for the appearance of massive stars; we adopt mixing via the Ledoux criterion as applied in MESA \cite{Jermyn2023}, and apply a mixing length  coefficient $\alpha_{\rm mlt}=2$. We allow step-function overshooting of 0.345 scale heights above the H-core burning region, which facilitates the growth of the He core. 

For late-stage burning, we allow locally super-Eddington regions of the convective envelope that form due to high opacity transitions in the cool outer envelope. We introduced additional energy transport efficiency in these regions, to avoid unrealistic inflation of stellar envelopes in the one-dimensional model. Previous three-dimensional models of the outer layers of convective massive stars have found large-scale convective plumes, time-variable turbulent layers, and effective porosity whereby radiation escapes through lower-density portions of envelope; these effects soften sharp features that occur in one-dimensional models \cite{2023Galax..11..105J,Jermyn2023}. We apply MESA's reduction factor to super-adiabatic temperature gradients that might otherwise occur [\cite{Jermyn2023}, their section 7.2]. The default treatment is active above $L(r)/L_{\rm Edd}(r) >0.5$, which would apply to the majority of our model stars' envelopes, and allows stellar cores to proceed in their evolution without impediment from the envelope. After experimenting with the MESA settings, we adopted a correction only to Eddington factor $L(r)/L_{\rm Edd}(r) >1.5$, which smooths the (unphysical) outer density inversion with little effect on the remainder of the envelope. Numerically, we adopt a convective Eddington parameter $\Gamma_{\rm c}=1.5$, the growth and damping terms for the time-dependent convection $\alpha_1=\alpha_2=20$, and a superadiabicity threhold $\delta_{\rm c}=0.1$ [\cite{Jermyn2023}, their equation 73].

At core helium exhaustion, our model star has a total mass of $10.81M_\odot$, of which $4.73 M_\odot$ are a core of helium and heavier elements. The star has a luminosity of $7.1\times10^4 L_\odot$ and an effective temperature of $3870$~K. From this epoch onward, the remaining evolution to core carbon depletion lasts $\sim2.5\times 10^4$~yr. To this late-stage burning, we manually apply a mass loss rate of $ 2.376\times10^{-4} M_\odot$~yr$^{-1}$. This mass loss rate is chosen for approximate consistency with the observed progenitor reddening, and to reach core carbon depletion at an effective temperature that is consistent with the $\sim 4500$~K progenitor in 2005-2012.  

The star reaches carbon core depletion (approximated by core carbon mass fraction $< 10^{-4}$) with a total mass of $5.02M_\odot$. Of this, $4.74 M_\odot$ is the helium core, and $0.28M_\odot$ is the hydrogen-rich envelope. The model has slightly enhanced surface helium abundance (by $\sim 10\%$ relative to its main sequence value), but retains approximately  solar surface metal mass fraction. The model star has an effective temperature of 4527~K, and a luminosity of $1.07\times 10^5L_\odot$, similar to the constraints on the progenitor star (Figure \ref{fig:prog} and Table \ref{tab:dusty_fits}). The model star's radius is $503R_\odot$, and its escape velocity is $\sim 60$~km~s$^{-1}$. The interior structure of the hydrogen envelope of this model is shown in Figure \ref{fig:profiles}.

\paragraph{\it Hydrogen-rich model:} 

To produce a hydrogen-rich counterpart to our hydrogen-poor model, we continue the evolution of the same model star beyond helium core exhaustion, with the same wind settings. At core carbon exhaustion, the mass loss rate is $4.2\times10^{-6}M_\odot$~yr$^{-1}$, substantially lower than in the hydrogen-poor model. The star is $10.80M_\odot$, of which $4.74 M_\odot$ are a core of helium and heavier elements. The star has a luminosity of $1.12\times10^5 L_\odot$ and an effective temperature of $3717$~K. The star's radius is 779$R_\odot$, and its escape velocity is 73~km~s$^{-1}$. This model is compared with the hydrogen-poor model in Figure \ref{fig:profiles}, which shows that the hydrogen-rich star has a more extended convection zone, and thus carries more of its (overall greater) mass at large radii.

\subsubsection*{Simulations of energy injection and fallback}

We adopt a simplified model of energy injection into these stellar envelopes by an outgoing shock, launched following the collapse of the core. The physics of shock energy deposition in failed SNe depends on factors including: the outward shock propagation through the infalling stellar material \cite{Coughlin2018,2018ApJ...863..158C}, the time-dependence of collapse of the core into a proto-neutron star \cite{Fernandez2018}, and on possible flow reversal and feedback from innefficient accretion onto a forming compact object\cite{2014MNRAS.439.4011G,2016ApJ...827...40G}\cite{Antoni2022,Antoni2023}. For qualitative comparison, we examine a model in which a given kinetic energy $E_{\rm sh}$ is added to stellar layers following a simple, power-law prescription \cite{Quataert2012}:
\begin{equation}
    v_{\rm sh} = v_0 \left( r \over R_*\right)^{\alpha_{\rm sh}},
\end{equation}
 where $v_{\rm sh}$ is the post-shock velocity, $v_0$ is a velocity scale, $r$ is the radius within the star, $R_*$ is the total stellar radius, and $\alpha_{\rm sh}$ is a power-law index. We assume $\alpha_{\rm sh}=1$. We integrate this profile over the stellar envelope to obtain the energy input,
\begin{equation}
    E_{\rm sh} = \int_0^{R_*} 2\pi r^2 \rho v_{\rm sh}^2 dr,
\end{equation}
and treat $v_0$ as a free parameter to match $E_{\rm sh}$ which we are trying to simulate. A fluid parcel is initially launched outward but may reverse and fall back if $v_{\rm sh}$ is less than the local escape velocity. Material expelled with $v_{\rm sh}<v_{\rm esc}$, where $v_{\rm esc}(r) = (2Gm/r)^{1/2}$ is the escape velocity, has a maximal radius
\begin{equation}
r_{\rm tr}(r) = r \left( 1- \left( {v_{\rm sh}(r) \over v_{\rm esc}(r) } \right)^2 \right)^{-1}
\end{equation}
where $r$ is the initial radius and $r_{\rm tr}(r)$ is the maximal, or ``turning" radius of material initially at radius $r$. 
The fallback time $t_{\rm fb} (r)$ for a shell of bound stellar material expelled to turning radius $r_{\rm rt}$ is then
\begin{equation}
    t_{\rm fb}(r) =  \left( 1 + {v_{\rm sh}(r) \over v_{\rm esc}(r) } \right) t_{\rm ff} (r_{\rm rt}(r)) 
\end{equation}
where $G$ is the gravitational constant, $m$ is the mass enclosed within radius $r$, and 
\begin{equation}
    t_{\rm ff} (r) = \pi \left(r^3 \over 8Gm\right)^{1/2}. 
\end{equation} 
If $v_{\rm sh}=0$, then $r_{\rm rt}(r) =r$ and the fallback time reduces to $t_{\rm ff}(r)$. The mass fallback rate is then $\dot M = d m_{\rm shell}/dt_{\rm fb}$ where $d m_{\rm shell}$ is the mass within a differential thin shell at radius $r$. At late times, when marginally bound material falls back, the asymptotic fallback rate changes with time $t$ as 
\begin{equation}
    \dot M \approx {2 \pi \over 3} r_0^2 \rho(r_0) v_{\rm esc}(r_0)  \left( {t \over t_{\rm ff}(r_0) } \right)^{-5/3},
\end{equation}
where $r_0$ and the quantities that depend on it are evaluated at the critical radius where the shock velocity equals the escape velocity $v_{\rm sh}(r_0) = v_{\rm esc}(r_0)$. The unbound mass $m_{\rm ej}$ is material with $v_{\rm sh}(r_0) > v_{\rm esc}(r_0)$, or 
\begin{equation}
    m_{\rm ej} = \int_{r_0}^{R_*} 4\pi r^2 \rho dr. 
\end{equation}

Random angular momentum in turbulently convective envelopes of supergiant stars cause substantial departures from spherical symmetry during collapse\cite{2014MNRAS.439.4011G,Quataert2019,Antoni2022}.
We expect individual plumes of stellar material to have a broad, isotropic specific angular momentum distribution with characteristic amplitude $j_{\rm plumes}$ of
\begin{equation}
    j_{\rm plumes} \approx r v_{\rm conv}
\end{equation}
where $v_{\rm conv}$ is the typical convective velocity at radius $r$. We determine its value from mixing length theory applied to the stellar models. These plumes, if they freefall, become rotationally supported at a radius $r_{\rm circ}$ 
\begin{equation}
    r_{\rm circ} \approx {j_{\rm plumes}^2 \over GM}.
\end{equation}
The specific angular momentum for material orbiting at the ISCO $j_{\rm ISCO}$ is approximately (depending on the BH's spin)
\begin{equation}
    j_{\rm ISCO} \approx {2 G M_{\rm BH} \over c}.
\end{equation}
where $c$ is the speed of light. The ratio of circularization radius to the radius of the innermost stable circular orbit $r_{\rm ISCO}$ is
\begin{equation}
    {r_{\rm circ} \over r_{\rm ISCO}} \approx \left( j_{\rm plumes} \over j_{\rm ISCO} \right)^2.
\end{equation}
Below this radius angular momentum is dynamically important and imparts turbulence.  The mass that falls inward is a function of radius, with turbulent convective motion preventing some degree of infall. We parameterize the accretion rate through radius $r$ as
\begin{equation}
    \dot m \approx \dot m(r_{\rm circ}) \left( r \over r_{\rm circ} \right)^\beta
\end{equation}
where $\beta \sim 0.5$ is a power-law index characterizing the suppression of mass accretion due to the turbulent motion, $\dot{m}(r_{\rm circ})$ is the rate that mass reaches the circularization radius as material falls in from the stellar envelope. Previous generalized simulations of turbulent infall \cite{Xu2023} derived $\beta \approx 0.7$ for gas with polytropic index $\gamma=5/3$, or $\beta \approx 0.6$ for RSG collapse \cite{Antoni2022}. We adopt the latter value, so
\begin{equation}
    \dot m_{\rm BH} \approx \dot m(r_{\rm circ}) \left( r_{\rm ISCO} \over r_{\rm circ} \right)^{0.6}.
\end{equation}
To compute the emergent luminosity from the accretion process in the late-time evolution, we adopt a simple radiative efficiency $\eta = 0.05$, the conversion factor from rest mass energy of the accreted material to the radiated luminosity $L = \eta \dot{m} c^2$.

\begin{figure}[!t]
    \centering
    \includegraphics[width=\textwidth]{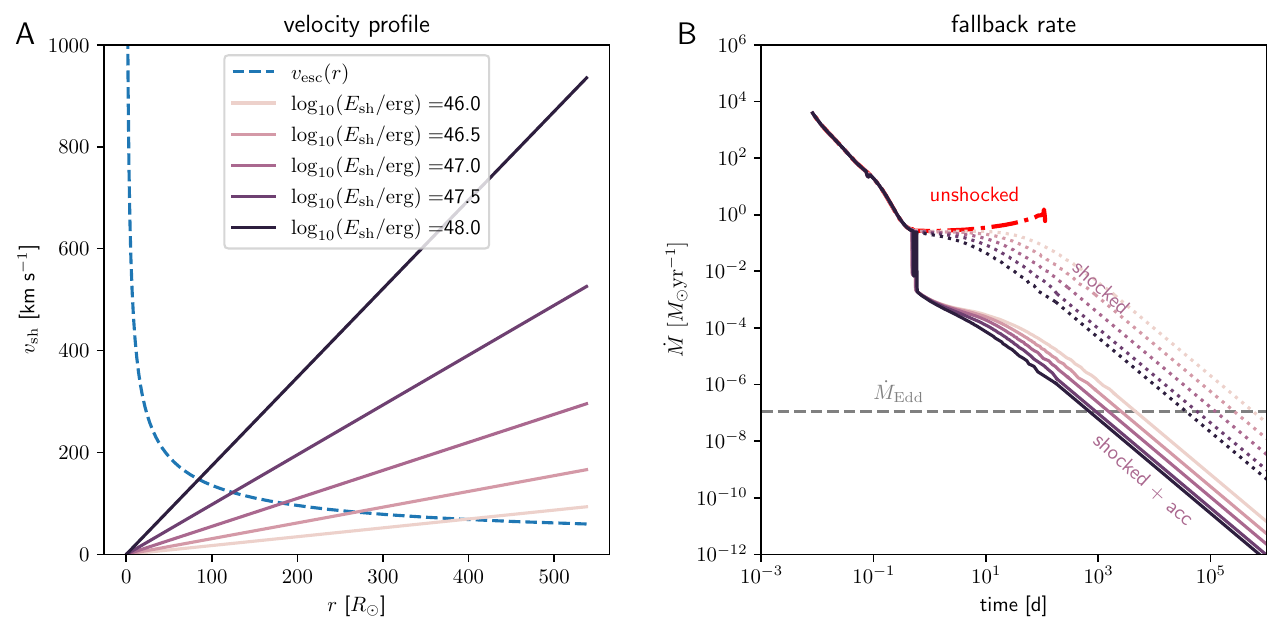}
    \linespread{1.0}\selectfont{}
    \caption{{\bf Radial velocity profiles of the shocked material and the resulting fallback accretion rate in the failed SN model}. (A) Colored lines show the radial velocity ($v_{\rm sh}$) profile of the shocked material for different shock energies $E_{\rm sh}$. The blue dashed line shows the radial profile of the local escape velocity $v_{\rm esc} (r)$; regions with $v_{\rm sh}>v_{\rm esc}(r)$ yield unbound material, while regions with $v_{\rm sh}<v_{\rm esc}(r)$ fall back towards the core. (B) Mass fallback rate (shown as colored lines corresponding to the same shock energies in panel A) due to the infall of the shocked material. The dotted lines show the fallback rate without accounting for the inefficient accretion induced by the angular momentum barrier of the initial convective envelope; solid lines show the same after implementing the suppression of mass accretion. The case of unshocked material ($E_{\rm sh} = 0$) fallback is shown as a red dot-dashed line. }
    \label{fig:fallbackmodel}
\end{figure}

Figure \ref{fig:fallbackmodel} shows the shock-injection, fallback, and accretion model applied to the H-poor progenitor MESA model, for a range of typical shock energies \cite{Fernandez2018}. Sufficiently strong shocks generate unbound material, and spread gas into marginally-bound trajectories that fall back at late-time with a dependence that asymptotes to $t^{-5/3}$. The effect of angular momentum is to greatly reduce the accretion efficiency, such that the BH accretion rate drops below its Eddington limit after about $10^3$~d post-collapse, for shock energies in the range of $10^{46}$ to $10^{48}$~erg. Compared to the case of an unshocked star (i.e. fallback without energy injection) which complete falls back within $\sim 100$\,d (Figure \ref{fig:fallbackmodel}B), the bulk of the mass ejection and fallback occurs thousands of days after core collapse.

\begin{figure}[!t]
    \centering
    \includegraphics[width=0.49\textwidth]{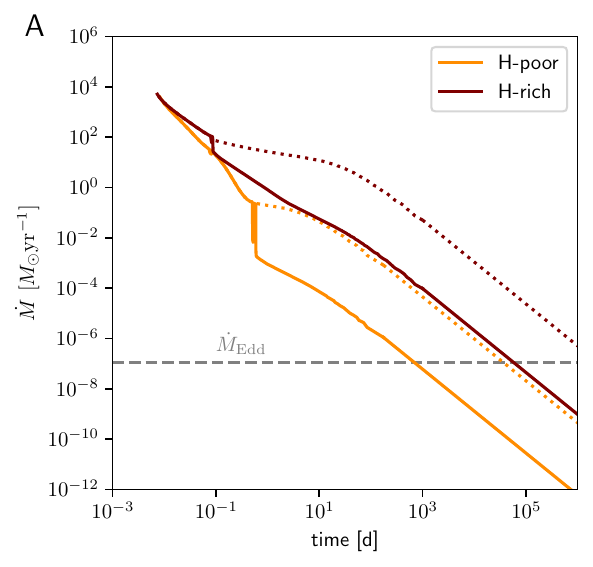}
    \includegraphics[width=0.49\textwidth]{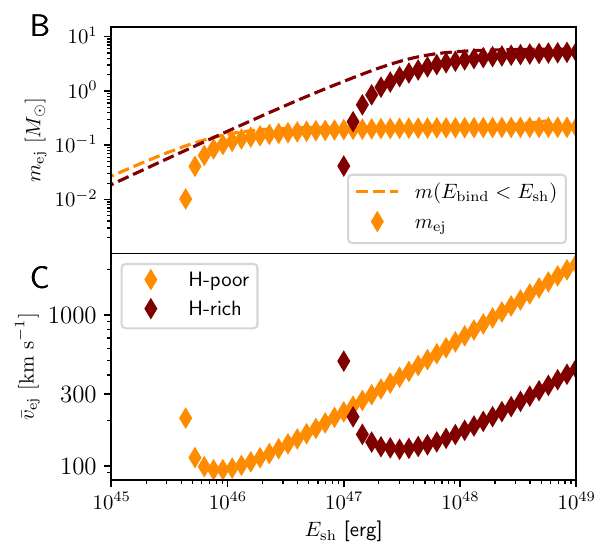}
    \linespread{1.0}\selectfont{}
    \caption{{\bf Comparison of fallback and mass ejection in H-rich and H-poor progenitor models.} (A) Same as Figure \ref{fig:fallbackmodel}B but comparing the fallback in the H-rich and H-poor models, with solid and dotted lines similarly showing the mass fallback rate with and without inefficient accretion. The Eddington mass accretion rate for a $5\,M_\odot$ compact object is shown as a gray horizontal dashed line for comparison. (B) Comparison of the ejected mass in the H-poor and H-rich models as function of the shock energy (diamonds). The dashed lines show the amount of mass with binding energy $E_{\rm bind}$ smaller than the input shock energy, with ejected mass asymptotically approaching this value for large shock energies. (C) Same as panel (B) but showing the velocity of the ejected material as a function of the shock energy.}
    \label{fig:unbound}
\end{figure}

Figure \ref{fig:unbound} compares the fallback and mass ejection for the H-rich and H-poor progenitor models. In the H-rich case, the late-time fallback from the mass-rich envelope is high (above the Eddington limit for $10^5$~d$\sim 300$~yr). The centrally concentrated H-poor model mostly accretes early, and drops below the Eddington limit in a few years. The expected ejecta properties also diverge. The H-rich model yields several solar masses of ejecta and transients of $L>10^6 L_\odot$ for hundreds of days. The H-poor model ejects about $0.1 M_\odot$ at a higher velocity of several hundred km s$^{-1}$. Figure \ref{fig:model} shows the predicted luminosity and timescale for the resulting outbursts (for H-recombination powered events). We find that the predicted transient from a hydrogen-poor supergiant would last tens of days, which is sufficiently brief to have been missed by the optical photometry (Figure \ref{fig:cadence_lim}). The faint luminosity of the predicted outbursts for low shock energies ($\sim 10^{39}$\,erg\,s$^{-1}$ for $E_{\rm sh} \lesssim 10^{48}$\,erg; Figure \ref{fig:model}) compared to the initial stellar luminosity ($\sim 5\times10^{38}$\,erg\,s$^{-1}$ from 2005-2012; Figure \ref{fig:prog}) would prevent the detection of an outburst given its irregular optical variability ($\approx 5\times$ in $r$-band; Figure \ref{fig:optirlc}) and consistent with the $10\sigma$ upper limits on any optical outburst (Figure \ref{fig:cadence_lim}).

\subsubsection*{Physical limits from X-ray non-detection}

The accretion of the turbulent convective envelope onto the BH is inefficient, causing most of the outer envelope material to be expelled (Figure \ref{fig:fallbackmodel}B; \cite{Antoni2023}). Both ejected and bound material can obscure emission from near the BH. We estimate the column density of hydrogen from the unbound component $N_{\rm H, ej}$ with $v_{\rm sh}>v_{\rm esc}$, which we assume to continue to expand at constant velocity, such that its radius is $v_{\rm sh}t$. Each radial zone in the initial stellar structure has mass $dm$. Then,
\begin{equation}
    N_{\rm H, ej} \approx {1\over m_p} \int \frac{dm}{4\pi (v_{\rm sh}t)^2 }
\end{equation}
where $m_p$ is the proton mass. $N_{\rm H, ej}$ scales as $t^{-2}$. The initially bound component that falls back toward the BH contains both the small fraction of material that accretes and the majority that is expelled instead. We neglect feedback from the passing shock\cite{Antoni2023} to estimate the column density of un-accreted material as a function of time. Most of the infall reverses around a scale of $r_{\rm circ}$, and expands from there. We estimate that the fallback driven outflow produces a column density $N_{\rm H, wind}$
\begin{equation}
    N_{\rm H, wind} \sim {1\over m_p} \frac{\dot m (r_{\rm circ})}{4\pi r_{\rm circ} v_{\rm esc}(r_{\rm circ})} 
\end{equation}
This component scales with the fallback rate, and therefore decays as $t^{-5/3}$ at late times. Thus, the ejecta column dissipates first, so at late times the remaining column density depends on the inefficient accretion-and-outflow process. 

\begin{figure}[!t]
    \centering
    \includegraphics[width=0.49\textwidth]{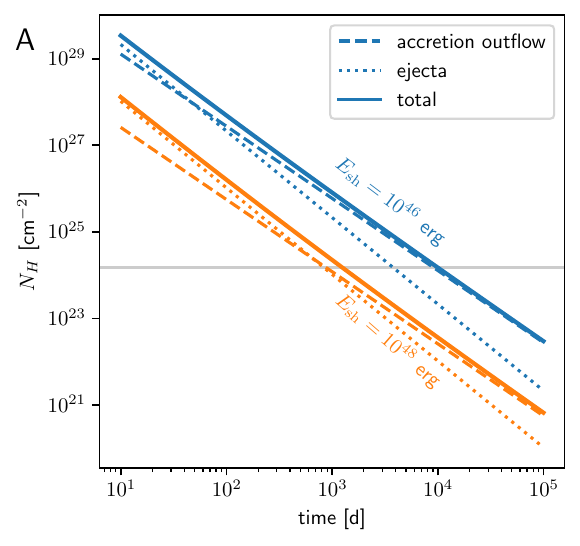}
    \includegraphics[width=0.49\textwidth]{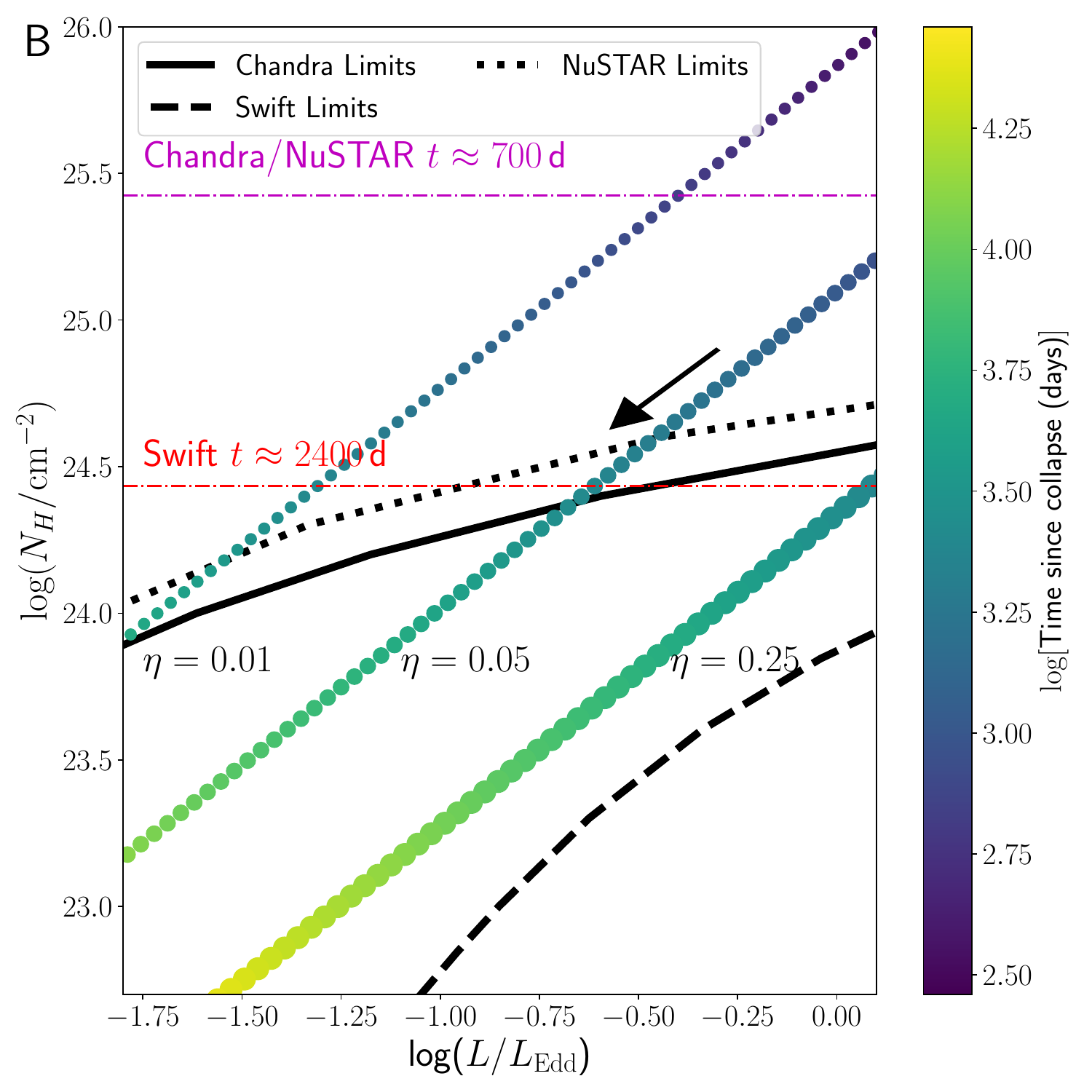}
    \linespread{1.0}\selectfont{}
    \caption{{\bf Predicted evolution of the hydrogen column density $N_H$ surrounding a newborn BH as a function of time and accretion luminosity ($L$), for different shock energies.} (A) Temporal evolution of the column density from the expanding ejecta (dotted lines), the accretion outflow from fallback material (dashed lines), and the total column density (solid lines) for two shock energies. The gray horizontal line shows $N_H \lesssim 1.5\times10^{24}$\,cm$^{-2}$, where the Thompson optical depth becomes $<1$ and the source may emerge in X-ray emission. (B)  Evolution of the column density as a function of the accretion luminosity (normalized to the Eddington luminosity of a $5$\,\Msun BH) for $E_{sh} = 10^{47}$\,erg, with the symbol colors indicating the time evolution (shown on the right color bar, and the arrow indicating the direction of increasing time) for three different radiative efficiencies ($\eta = 0.01, 0.05, 0.25$). The magenta dot-dashed and red dashed lines show the estimated $N_H$ from the model at the epoch of the archival Chandra/NuSTAR (end of 2015) and Swift X-ray observations, respectively. The black solid, dotted and dashed lines show the detection threshold for X-ray emission for the respective Chandra, NuSTAR and Swift X-ray observations, where only regions below the lines would be detectable in the data.}
    \label{fig:nh}
\end{figure}

Figure \ref{fig:nh} estimates the column density as a function of time from our model. This includes both the unbound and reversing-fallback components. At late times, the more uncertain accretion outflow component dominates, due to its shallower time decay. For characteristic energies of $\sim 10^{47}$\,erg, the column density along the line of sight drops below $10^{24}$\,cm$^{-2}$ -- the approximate level to allow soft X-rays to penetrate -- from a few $\times 10^3$ to $10^4$~d. To quantitatively compare against our observations, we track the expected evolution of our model as a function of the surrounding column density and accretion luminosity (both of which decrease rapidly with time). Figure \ref{fig:nh} compares those model predictions to the estimated total $N_H$ at the epochs of the archival Chandra, NuSTAR and Swift X-ray observations and the phase space of $N_H$ and $L$ that are ruled out by the X-ray non-detections (assuming a $\Gamma = 2$ intrinsic X-ray spectrum). The calculations indicate that the source was too heavily obscured to be detected at the epochs of the archival X-ray observations.


\subsection*{Supplementary Text}

\subsubsection*{The case of NGC\,6946-BH1}

M31-2014-DS1 shares several similarities with a previously reported  failed SN candidate NGC\,6946-BH1\cite{Gerke2015, Adams2017, Basinger2021}. Both have a luminous supergiant progenitor, bolometric decay and a remnant dominated by infrared emission. While the progenitor was interpreted as a RSG based on a comparison of its colors to single star evolutionary tracks\cite{Adams2017}, its pre-outburst colors were shown to be consistent with a hydrogen depleted yellow supergiant \cite{Humphreys2019}. The progenitor of NGC\,6946-BH1 also had an infrared excess \cite{Adams2017} indicating of intense terminal mass loss. For quantitative comparisons, we repeated the same analysis as above but applied to NGC\,6946-BH1 by producing a stellar model that matches the reported luminosity and temperature of its progenitor ($\log (L/L_\odot) = 5.29^{+0.04}_{-0.06}$, $T_{\rm eff} = 4480^{+1670}_{-320}$; Figure \ref{fig:prog}B).

Adopting the same MESA parameters, we model the NGC 6946-BH1 progenitor as an initially $17.5\,M_\odot$ star with solar metallicity. We evolve this star to helium core depletion with the same mass loss prescription, at which point it has a mass of $12.62M_\odot$, of which $6.89M_\odot$ are a core of helium and heavier elements. To produce a hydrogen-poor model consistent with the yellow-supergiant hypothesis, after helium depletion, we apply an enhanced mass loss rate of $3.915\times10^{-4}M_\odot$~yr$^{-1}$ until carbon is depleted in the star's core. At that point, the model star has a total mass of $7.52M_\odot$, of which $6.90M_\odot$ are the core and $0.62M_\odot$ are hydrogen-rich. The surface helium abundance by mass is slightly enhanced relative to its initial state (0.31 versus 0.25 in the initial model), while the surface metal abundance remains solar.  The model star has a luminosity of $1.81\times10^{5}L_\odot$, and an effective temperature of $4562$~K.

\begin{figure}[!t]
    \centering
    \includegraphics[width=0.49\linewidth]{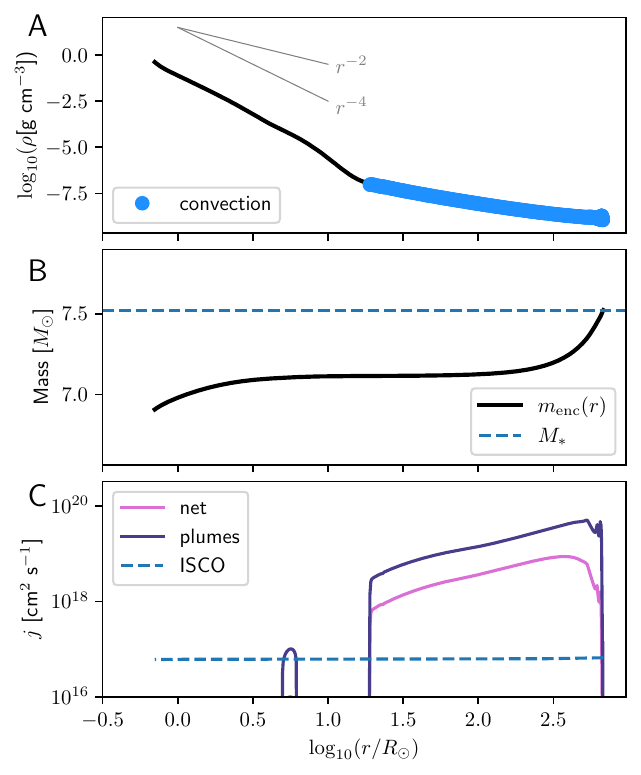}
    \includegraphics[width=0.49\linewidth]{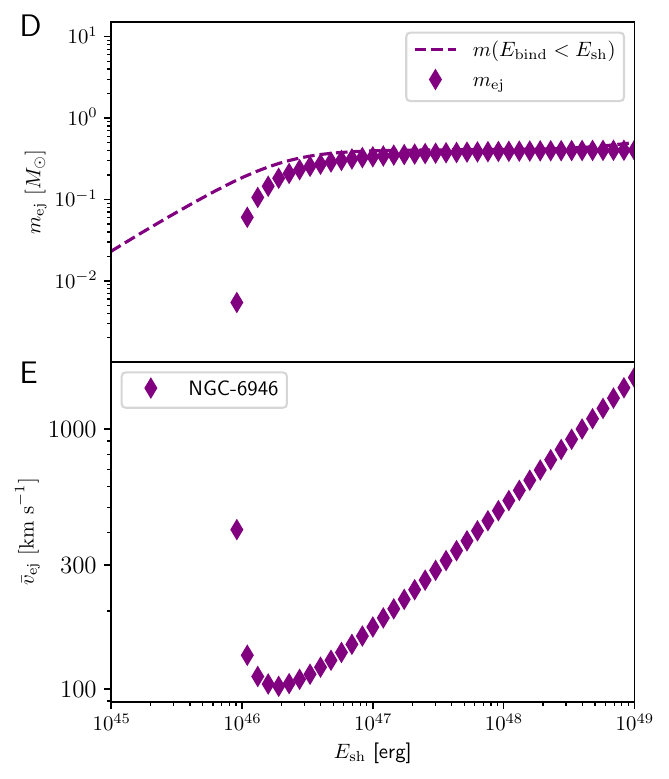}
    \linespread{1.0}\selectfont{}
    \caption{{\bf Stellar model and predicted outburst properties for the progenitor of NGC\,6946-BH1}. (A-C) Same as Figure \ref{fig:profiles}A-C for the NGC\,6946-BH1 terminal stellar model. (b) Same as Figure \ref{fig:unbound}B-C, but for the NGC\,6946-BH1 stellar model.}
    \label{fig:ngc6946_profile}
\end{figure}

\begin{figure}[!t]
    \centering
    \includegraphics[width=0.49\linewidth]{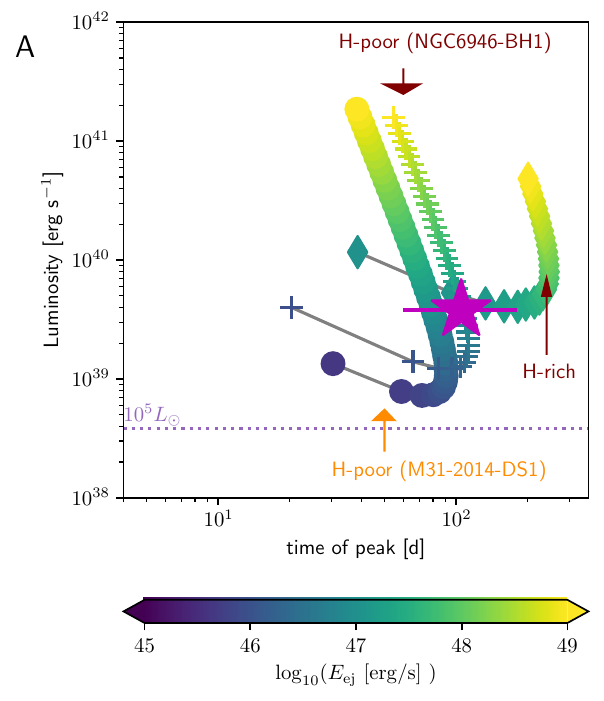}
    \includegraphics[width=0.49\linewidth]{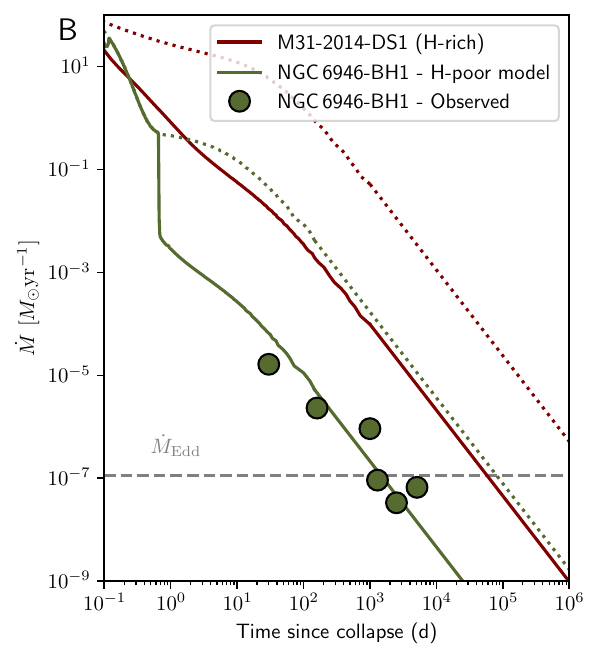}
    \linespread{1.0}\selectfont{}
    \caption{{\bf The predicted outburst properties and late-time mass fallback evolution for the model for NGC\,6946-BH1, compared to the H-poor and H-rich models for M31-2014-DS1.} (A) Same as Figure \ref{fig:model}A for the H-poor model for M31-2014-DS1 (circles), the H-rich model for M31-2014-DS1 (diamonds) and the progenitor model for NGC\,6946-BH1 (plus symbols) with the shock energy shown on the color bar. The magenta star shows the observed luminosity and duration of the NGC\,6946-BH1 outburst. The horizontal dotted line shows progenitor luminosity for M31-2014-DS1. (B) Same as Figure \ref{fig:fallbackmodel}B but comparing the mass fallback evolution for the H-rich progenitor model of M31-2014-DS1 and the H-poor progenitor model of NGC\,6946-BH1, for shock energy of $10^{48}$~erg. The observed bolometric light curve of NGC\,6946-BH1 is shown after converting luminosity into mass accretion rate assuming a radiative efficiency of $\eta = 0.05$.}
    \label{fig:ngc6946_outburst}
\end{figure}

Figure \ref{fig:ngc6946_profile} shows that the higher hydrogen envelope mass and larger radius for this stellar model (compared to M31-2014-DS1) results in the ejection of more mass ($\sim 0.3$\,\Msun) at higher velocities ($\sim 300$\,km\,s$^{-1}$).  The outburst properties associated with H-recombination for the ejecta are shown in Figure \ref{fig:ngc6946_outburst}, compared to the H-poor and H-rich progenitor models for M31-2014-DS1. We find that the larger ejecta mass produces a longer duration outburst in the H-poor models. For NGC\,6946-BH1, the total duration of the outburst was poorly constrained as between 3 and 11 months\cite{Adams2017}. Taking the rise time of the outburst to be half of the total duration, Figure \ref{fig:ngc6946_outburst}A shows that the observed outburst properties of the source are consistent with a $\sim 10^{47}$\,erg shock powered by neutrino mass loss.

Figure \ref{fig:ngc6946_outburst}B shows the evolution of the mass fallback rate as the loosely bound material falls back into the stellar core. The $\approx 3\times$ more massive H-envelope leads to a longer time of $\approx 2000$\,d before the accretion falls below the Eddington rate, then fades by a factor of 10 over a few thousand days after the collapse. NGC\,6946-BH1 was previously reported to fade to $\approx 15$\% of its progenitor luminosity in $\sim 3000$\,d \cite{Adams2017}, Figure \ref{fig:ngc6946_outburst}B compares its bolometric light curve (converted to effective mass accretion rate) to our model. Observations of the NGC\,6946-BH1 remnant at $\approx 5100$\,d show that it has a luminosity of $\sim 10-20$\% of the progenitor\cite{Kochanek2023, Beasor2023}. Earlier epochs of observation did not have mid-IR coverage (therefore the bolometric luminosity was poorly constrained). Nevertheless, the slower fading of NGC\,6946-BH1 is consistent with our model, which predicts fading by a factor of 2 to 3 over a decade.

\subsection*{Comparison to rates of failed SNe}

An observational strategy has been previously proposed \cite{Kochanek2008} to search for the disappearance of massive stars due to stellar collapse into a BH, possibly accompanied by a low luminosity transient. Given the typical lifetime of an evolved supergiant ($\sim 10^6$\,years), a survey monitoring $\sim 10^6$ supergiants could identify massive stellar death (either as a successful or failed SN) at a rate of $\approx 1$\,yr$^{-1}$ \cite{Kochanek2008}, and constrain the fraction of failed SNe. A $\approx 11$\,yr survey \cite{Gerke2015, Adams2017b, Neustadt2021} led to the identification of NGC\,6946-BH1 \cite{Adams2017} and another less constrained candidate \cite{Neustadt2021}. A previous search of archival data for low luminosity optical transients reported no detections or candidates \cite{Byrne2022}.

Our NEOWISE search was sensitive to variability of supergiants in M31 and M33 down to a luminosity threshold of $\gtrsim 10^{37}$\,erg\,s$^{-1}$. Taking the luminosity of the MIR brightening in M31-2014-DS1 ($\approx 10^{38}$\,erg\,s$^{-1}$), and considering that dust luminosity scales linearly with the ejected mass in the optically thin brightening phase, our search was sensitive to all outbursts with ejected mass $\gtrsim 10^{-2}$\,\Msun ($\approx 10$\% of that M31-2014-DS1). This covers the entire range of expected H-envelope masses in supergiant stars \cite{Yoon2017}, the survey included $\approx 2 \times 10^4$ supergiants (with optical luminosity $\gtrsim 3.5 \times 10^3$\,L$_\odot$) in M31 and M33 [\cite{Kochanek2008}, their Figure 4] and had a $\approx 15$\,yr baseline.

The fraction of massive star deaths that result in failed SNe ($f$) is unknown, with previous studies proposing a lower limit of $f \gtrsim 0.25$ based on the local BH density \cite{Kochanek2008}, estimate $f \approx 0.04 - 0.4$ based on the previous survey \cite{Neustadt2021}, and an upper limit of $f < 0.61$ \cite{Byrne2022}. Assuming a typical supergiant lifetime of $\sim (5 - 10) \times 10^5$\,yr, we therefore estimate the number of such events in M31 and M33 to be $\approx 0.01 - 0.2$ for $f \approx 0.05 - 0.5$ over a duration of $\approx 15$\,yr. Therefore, our identification of one such event is not improbable but fortuitous, even if the fraction of failed events is close to the upper limit. A similar calculation for the previous survey that monitored $\sim 10^6$ supergiants over the same duration \cite{Kochanek2008} found an expected number of $0.5 - 5$ events, consistent with their results (two candidates). The high inferred fraction is consistent with theoretical predictions that BH formation can occur in stars with masses as low as $\approx 13$\,\Msun.




\begin{table}[!h]
    \centering
    \caption{Archival space-based photometry of M31-2014-DS1 from HST and SST. Uncertainties are $1\sigma$ confidence, and upper limits are $5\sigma$.}
    \begin{tabular}{lccc}
    \hline
    Epoch (MJD) & Filter & AB Mag & Instrument \\
    \hline
    \hline
56115 & F110W & $16.86 \pm 0.05$ & HST/WFC3-IR \\
56115 & F160W & $16.83 \pm 0.05$ & HST/WFC3-IR  \\
56273 & F475W & $19.15 \pm 0.05$ & HST/ACS \\
56273 & F814W & $17.05 \pm 0.10$ & HST/ACS \\
56115 & F336W & $23.25 \pm 0.10$ & HST/WFC3-UVIS  \\
56115 & F275W & $>26.18$ & HST/WFC3-UVIS  \\
53391 & Channel\,1 ($3.6$\,\textmu m) & $16.37 \pm 0.01$ &  SST/IRAC\\
53391 & Channel\,2 ($4.5$\,\textmu m) & $16.47 \pm 0.01$ &  SST/IRAC\\
53391 & Channel\,3 ($5.8$\,\textmu m) & $16.05 \pm 0.01$ &  SST/IRAC\\
53391 & Channel\,4 ($8.0$\,\textmu m) & $15.95 \pm 0.01$ &  SST/IRAC\\
53243 & Channel\,1 ($24.0$\,\textmu m) & $15.69 \pm 0.10$ & SST/MIPS \\
59617 & F606W & $>28.48$ & HST/WFC3-UVIS\\
59617 & F814W & $27.30 \pm 0.23$ & HST/WFC3-UVIS\\
\hline
    \end{tabular}
    \label{tab:prog_photo}
\end{table}

\begin{table}
    \centering
    \small
    \caption{Best-fit DUSTY parameters for the progenitor and remnant of M31-2014-DS1.}
    \begin{tabular}{lccccccc}
    
    \hline
    Model & $F$ ($10^{-12}$\,erg\,cm$^{-2}$\,s$^{-1}$) & $\log (L/L_\odot)$ & $T_{\rm eff}$ (K) & $T_d$ (K) & $\tau$ & $Y$ & $A_V$ (mag)\\
    \hline
    \hline
    Progenitor & $5.0^{+1.0}_{-0.7}$ & $4.97^{+0.08}_{-0.07}$ & $4500^{+610}_{-520}$ &  $880^{+180}_{-140}$ &  $2.8^{+0.4}_{-0.5}$ &  $54^{+31}_{-32}$ &  $0.1^{+0.2}_{-0.1}$ \\
    Remnant & $0.5^{+0.2}_{-0.1}$ & $3.97^{+0.15}_{-0.10}$ & $6800^{+4100}_{-1800}$ & $960^{+220}_{-170}$ & $22.0^{+1.5}_{-1.5}$ & $10$ (fixed) & $0.1$ (fixed) \\
\hline
    \end{tabular}
    \label{tab:dusty_fits}
\end{table}


\clearpage 



\end{document}